# Autonomous Discovery of Tough Structures

**Authors**

Kelsey L. Snapp[1], Benjamin Verdier[2], Aldair Gongora[1], Samuel Silverman[2], Adedire D. Adesiji[1], Elise F. Morgan[1,3,4], Timothy J. Lawton[5], Emily Whiting[2], and Keith A. Brown[1,3,6]

1 Department of Mechanical Engineering, Boston University, Boston, MA, USA

2 Department of Computer Science, Boston University, Boston, MA, USA

3 Division of Materials Science & Engineering, Boston University, Boston, MA, USA

4 Department of Biomedical Engineering, Boston University, Boston, MA, USA

5 Soldier Protection Directorate, US Army Combat Capabilities Development Command Soldier Center, Natick, MA, USA

6 Physics Department, Boston University, Boston, MA, USA

**Abstract**

**A key feature of mechanical structures ranging from crumple zones in cars to padding in packaging is their ability to provide protection by absorbing mechanical energy.[1,2] Designing structures to efficiently meet these needs has profound implications on safety, weight, efficiency, and cost.[3,4] Despite the wide varieties of systems that must be protected, a unifying design principle is that protective structures should exhibit a high energy-absorbing efficiency, or that they should absorb as much energy as possible without mechanical stresses rising to levels that damage the system.[4,5] However, progress in increasing the efficiency of such structures has been slow due to the need to test using tedious and manual physical experiments. Here, we overcome this bottleneck through the use of a self-driving lab to perform >25,000 machine learning-guided experiments in a parameter space with at minimum trillions of possible designs. Through these experiments, we realized the highest mechanical energy absorbing efficiency recorded to date. Furthermore, these experiments uncover principles that can guide design for both elastic and plastic classes of materials by incorporating both geometry and material into a single model. This work shows the potential for sustained operation of self-driving labs with a strong human-machine collaboration.**

**Main Text**

Introduction

Structural motifs define the ways we efficiently use materials. For instance, the ubiquity of "I" beams in architecture is due to the efficiency of this shape in resisting both shear and bending.[6,7] Natural structures feature similar examples such as the hollow circular cross-section of bamboo providing high bending and torsional resistance.[8–11] For the large class of structures designed to provide protection under a compressive load, the key property to consider is the total mechanical energy absorbed during compression.[1,2,12,13] However, there are fundamental restrictions to absorbing energy, for example, that the stress must be held below a level that would damage the system to be protected. Therefore, it is useful to define an energy-absorbing efficiency $K_s$, a non-dimensional measure of how much energy is absorbed without surpassing a given threshold stress.[4,5] Unfortunately, $K_s$ is difficult to optimize because most of the energy absorbed by a structure designed for mechanical protection occurs beyond the elastic regime where deformations are highly non-linear, often feature dynamic self-self contacts, and are challenging to model.

As a result of the challenge of designing efficient absorbers of mechanical energy, much work has focused on known, relatively simple motifs such as honeycomb lattices or cylindrical shells that have an analytical basis for performing well.[4,14] Others have drawn inspiration from nature to identify more complex structural motifs.[15–18] Computational approaches including finite element analysis and machine learning-based approaches have also been widely used to design complex geometries.[19–25] These computational approaches pair well with additive manufacturing, which allows the fabrication of extremely complex designs.[26–30] Nevertheless, the fabrication of candidate structures is often the limiting step in the design process and is commonly limited to validating designs. Thus, improvements to $K_s$ remain slow: to date, additively manufactured structures designed for energy absorption typically feature $K_s < 50\%$ (Figure S1). There exist better synthetic materials, the best being a plastic foam reported to have reached $K_s = 68.1\%$.[31] However, this record is surpassed by nature: Balsa wood has the highest previously achieved $K_s$, 71.8%, showing the value of millions of years of evolution.[32] It is clear that new approaches are needed if the performance envelope of this important property is to be improved.

Here, we utilize a self-driving lab (SDL) to test >25,000 additively manufactured structures in a large-scale data-driven campaign to discover structures with superlative $K_s$. SDLs are robotic research systems that select, perform, and analyze physical experiments without needing human intervention,[33,34] and they have been productively employed in chemistry,[35,36] materials science,[37] mechanics,[38] and microscopy.[39,40] Motivated by the observations that SDLs can progress toward user-chosen goals faster than either high-throughput experimentation[41] or tests chosen by subject matter experts,[42] we hypothesized that an SDL allowed to explore seven polymers in an 11-dimensional parameter space over trillions of possible designs could discover new structural motifs that advance the frontier of $K_s$. The result of this sustained human-machine collaboration is that we realize a structure with $K_s = 75.2\%$, the highest value reported to date. In addition to showing the opportunities for SDLs to overcome design barriers, this campaign resulted in a vast, labeled dataset that has implications for both mechanics and design more generally. For instance, we explore two high-performing structural motifs and find that they exhibit consistent performance within classes of materials, namely plastic or hyperelastic polymers. Finally, aggregate analysis of this data provides general design heuristics that allow for the efficient selection of materials and structures.

Main Body

As a motivating example to explore the considerations that define and limit $K_s$, we consider the compressive behavior of a cylindrical shell composed of a hyperelastic thermoplastic polyurethane (TPU). When tested in compression, the resulting force $F$-displacement $D$ curve shows an initial elastic region, a yield point, and then complex post-yield behavior that originates from combinations of plastic deformation, buckling and other large elastic deformations, and reentrant contact (Figure 1a). To compute $K_s$, $F$-$D$ is first converted to stress $\sigma$ vs. strain $\varepsilon$ for the effective medium using the dimensions of the component (Figure S2). Defining $K_s$ requires specifying a threshold stress $\sigma_t$ that is typically associated with the strength of the system to be protected. Graphically, $K_s$ represents the amount of energy absorbed by the component while $\sigma \leq \sigma_t$ (Figure 1a – blue region) relative to the maximum energy that could be absorbed during complete compression ($\varepsilon = 1$) without exceeding $\sigma_t$ (Figure 1a – red rectangle). To compute $K_s$ at a specific $\sigma_t$, we numerically evaluate $K_s = \sigma_t^{-1} \int_0^{\varepsilon_t} \sigma(\varepsilon) d\varepsilon$, where $\varepsilon_t$ is the greatest strain at which $\sigma \leq \sigma_t$ for all $0 < \varepsilon \leq \varepsilon_t$. Interestingly, for most structures, $K_s(\sigma_t)$ has a single well-defined maximum $K_s^*$ at a characteristic threshold stress $\sigma_t^*$. In the example of Figure 1a, the cylindrical shell is limited to $K_s^* = 39.8\%$ due to significant post-yield softening. To maximize $K_s^*$, a flat post-yield region and a delay of densification until large $\varepsilon$ are both desirable. Unfortunately, this knowledge alone does not provide a prescription for how to adjust the structure to obtain these desired behaviors.

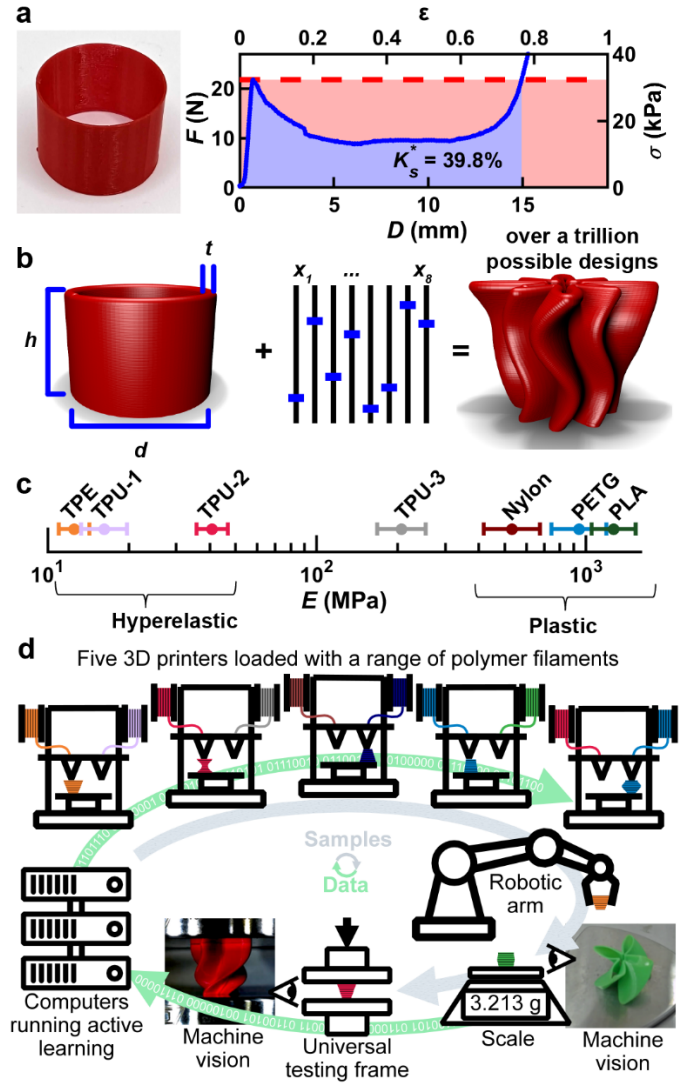

**Fig. 1 | Challenge of designing energy absorbing structures. a,** Force $F$ vs. displacement $D$ and effective medium stress $\sigma$ vs. compressive effective medium strain $\varepsilon$ for an additively manufactured cylindrical shell made of thermoplastic polyurethane (TPU). Maximum energy absorbing efficiency $K_s^*$ is calculated at an optimum threshold stress $\sigma_t^*$ (dashed line) by dividing the energy absorbed while $\sigma \leq \sigma_t^*$ (blue region) by the theoretical maximum amount absorbed (red rectangle). **b,** Eleven independent geometric parameters including diameter $d$, height $h$, wall thickness $t$, and eight other parameter $x_{1-8}$ that together define a generalized cylindrical shell (GCS). When combined, at least trillions of unique designs are possible. **c,** Modulus $E$ of the seven polymers studied as determined by compression tests. Error bars represent one standard deviation. **d,** Schematic showing an autonomous research system in which five 3D printers are used to fabricate polymeric structures that are automatically weighed, imaged, and tested using quasistatic compression. The output of this testing is automatically interpreted and used to select subsequent designs to test.

We hypothesized that programmed perturbations to the geometry of a cylindrical shell could tailor the complex post-yielding behavior to drastically increase $K_s^*$. While cylindrical shells are typically defined by a small number of geometric parameters, namely their diameter $d$, height $h$, and thickness $t$, we augmented these to form an 11-parameter family of structures termed generalized cylindrical shells (GCS) (Figure 1b). In addition to *t, h,* and *d*, a GCS is defined by eight additional parameters $x_{1-8}$ including four that adjust the cross-sections of the top and bottom of the shell, one to define the perimeter of the top relative to the perimeter of the bottom, and three to describe the rotation of the perturbations from top to bottom (Figure S3). Furthermore, because the GCS space does not have circular cross sections for most parts, we define $d = \frac{P}{2\pi}$, where *P* is the average perimeter for the part. We note that each of the eleven parameters would only need twelve unique values for their combinations to surpass a trillion unique designs. Given that each parameter is continuous and can be assigned many more than twelve values, we consider trillions of unique designs to be a lower limit to the size of the parameter space. However, since all the resulting structures are topologically equivalent to cylindrical shells, they can be fabricated using extrusion-based additive manufacturing by continuously extruding material, thus making this parameter space intrinsically designed for additive manufacturing. In addition to the geometric parameters, we sought to explore a variety of polymers. Therefore, we considered seven materials that included those that are hyperelastic such as a thermoplastic elastomer (TPE) and TPU, those that plastically deform such as polylactic acid (PLA), and materials that fall in between these two distinct classes (Figure 1c). These materials can be characterized based on their elastic modulus *E*, their plateau strength $\sigma_p$, and rebound strain (Figure S4).

      To efficiently search the effectively infinite GCS parameter space, we employed the Bayesian experimental autonomous researcher (BEAR), a customized SDL developed to combine additive manufacturing of polymers and mechanical testing (Figure 1d). The BEAR is a closed-loop system in which samples are printed using one of five fused filament fabrication (FFF) printers, automatically retrieved using a six-axis robotic arm, and then characterized using a scale, machine vision, and uniaxial compression testing. After testing, the information was automatically analyzed to determine whether the test was of acceptable quality. Subsequent experiments were selected using Bayesian optimization, which entailed conditioning a surrogate model of the mechanical performance using all previously measured GCS components and then selecting combinations of designs and materials that maximized a specified acquisition function. The SDL autonomously performed these tasks to choose, perform, and analyze experiments at a typical pace of ~50 experiments per day. Collectively, 25,387 experiments were performed using seven different materials, with 13,250 experiments resulting in valid data. This system is an evolution of an SDL developed by our research group.[41] A picture of the system (Figure S5) as well as full details on the hardware (Figure S7) and software (Figure S8) used as part of the BEAR are provided in the methods and supporting information.

      An extensive SDL campaign proceeded as a continuous human-machine collaboration where the responsibilities were shared between the SDL and the human team (Figure S9). Progress in the campaign can be visualized by tracking $K_s^*$ measured for each experiment along with a running maximum throughout the campaign (Figure 2a). The continuous progression was a result of both persistent experimentation by the SDL and choices made by the human team based on the progress of the SDL. Interestingly, large jumps in performance were typically either due to serendipity (*i.e.* the SDL chose a fortuitous region), or a human-led intervention. For example, we highlight three human interventions based on observing the progress of the SDL.

First, prior to experiment 4,829, the SDL was programmed to select experiments based on $K_s$ at one specific $\sigma_t$. However, we noted that there were several different reasons why a specific sample would have a low $K_s$, so we needed to provide the SDL with more information. We hypothesized that tracking both $K_s^*$ and $\sigma_t^*$ from each experiment would allow for more meaningful information to be extracted by the SDL. After implementing this change, the SDL rapidly increased from 45% to 55% $K_s^*$. As a second example, at experiment 15,678, we noted that a large fraction of plastic components were failing the height quality control check but passing the mass check. We had been heating the print bed after printing to facilitate the automated removal of components, but determined that the forces exerted during removal could deform the plastic components. Upon changing the SDL to cool plastic components prior to removal, the system proceeded to make a series of jumps from 60 to 68% $K_s^*$. Finally, at experiment 17,730 we noted that the predictive model used by the SDL was systematically underpredicting $K_s^*$ for high performing components, so we implemented a process where the proposed experiment was selected using a model built only on data close to the best observed experiment, a process similar to algorithms such as TURBO or ZOMBI.[43,44] This intervention led the SDL to progress from 70.6% to 75.2% $K_s^*$. A summary of significant human-led actions is provided in Figure S10.

    The culmination of these adjustments and continued experimentation by the SDL resulted in the observation of 75.2% $K_s^*$, a value that was higher than had been previously reported. The performance of this superlative experiment is shown in Figure 2b, which shows the $\sigma - \varepsilon$ curve and photographs of the component at different stages of compression. It is clear from the flatness of the post-yield region together with the photographs that the SDL has discovered a way for buckling and other large elastic deformations, plasticity, and reentrant contact to work together to achieve a remarkably flat plateau until densification initiated at ~80% strain. Interestingly, this component was composed of PLA, which is not commonly regarded as a high-performance material. Upon repeated experiments, the design, which we termed Palm, had an average $K_s^* = 73.1 \pm 0.9$. Although Palm printed in PLA had the largest single value of $K_s^*$, 75.2%, observed in the entire campaign, we discovered other components that had higher average $K_s^*$ values than Palm.

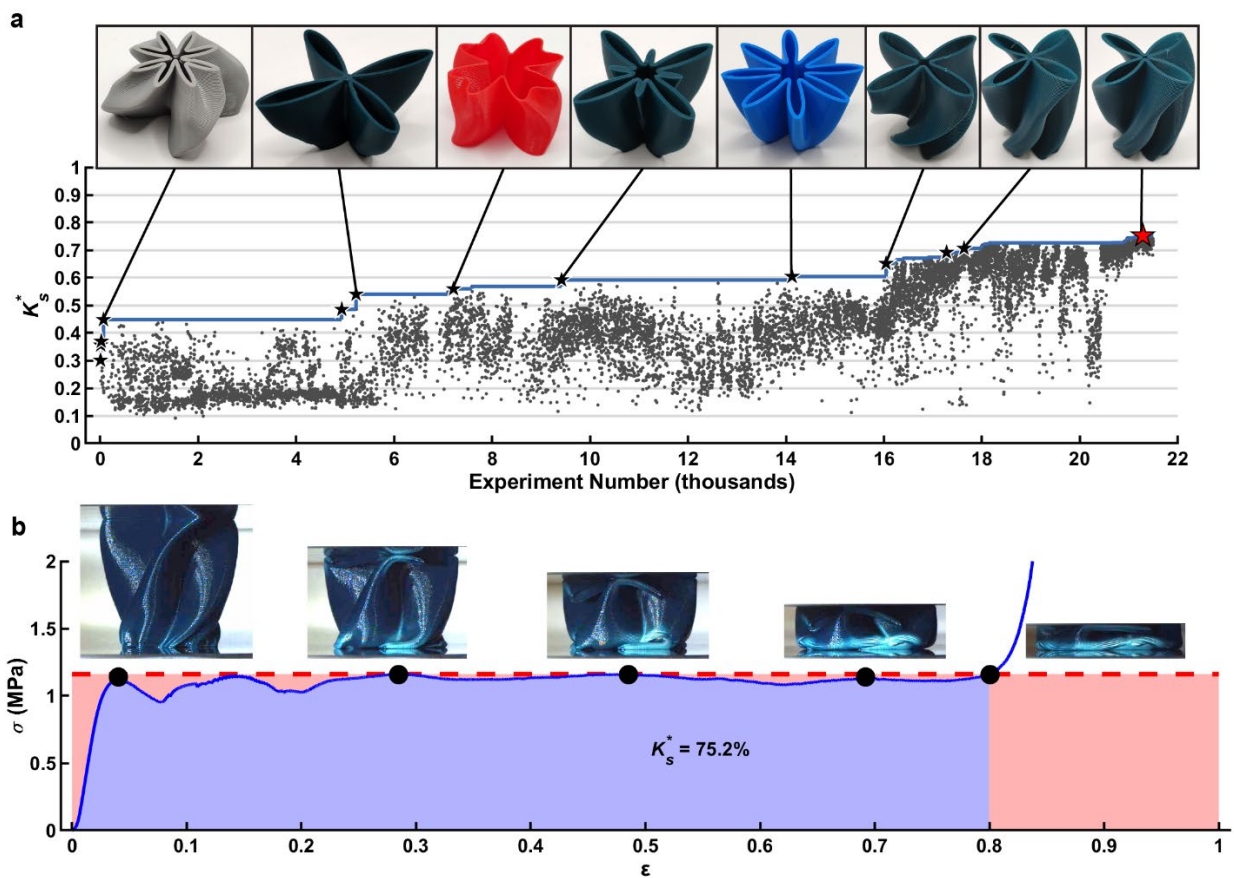

**Fig. 2 | Autonomous research campaign to find highly efficient structures. a,** Each $K_s^*$ measured over the first ~21,500 experiments performed. Pictures highlight noteworthy components (black stars) and the highest performing structure (red star). Larger versions of images are included as Figure S11. The color of the pictured components is indicative of the material used, with Green indicating PLA, Blue indicating PETG, and Red/Gray indicating different types of TPU. The solid blue line denotes the running best $K_s^*$ observed. **b,** $\sigma$ vs. $\varepsilon$ for experiment 21,285, named Palm, which resulted in $K_s^* = 75.2\%$. Inset photographs show the state of the component at various points indicated on the curve (images enhanced to improve clarity – originals given as Figure S12).

To explore variations in performance across different material classes, we selected two high-performing designs discovered in different materials. The design discovered in PLA with the highest average $K_s^*$, termed Willow, is tall and has a compact center region (Figure 3a). Testing 15 identically prepared samples of the Willow design resulted in a tight distribution of yield forces with variations in the post-yield plateau. Nevertheless, PLA components made using this design exhibit $K_s^* = 73.3 \pm 0.9\%$, showing a consistent performance above previously reported maxima. The highest performing design discovered for TPU-2 is termed Iroko and consists of a relatively short and open design (Figure 3b). We observed more substantial variations among the 15 Iroko $\sigma - \varepsilon$ profiles, and the average $K_s^* = 53 \pm 4\%$ was substantially lower than that of the best plastic components. The differences in attainable $K_s^*$ between PLA and TPU-2 can be explained by considering that these are different material classes, with PLA being a glassy polymer that exhibits substantial plastic deformation while TPU-2 is a hyperelastic elastomer. This difference in properties is most evident in their post compression behavior in which the TPU component recovers ~99% of its height one minute after compression while the PLA component is permanently flattened to ~23% of its initial height.

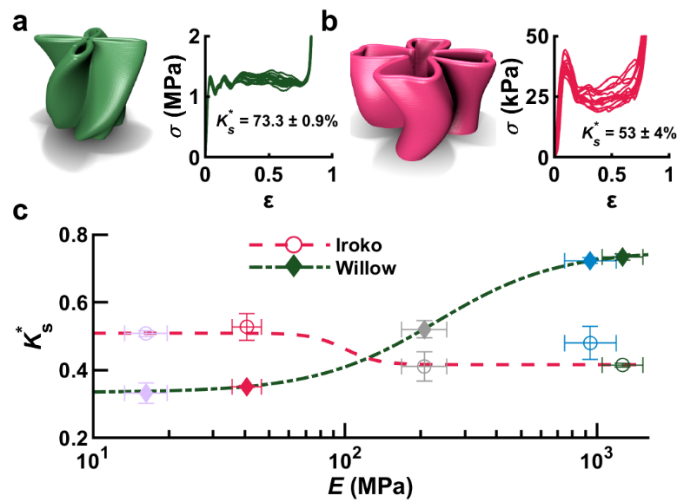

**Fig. 3 | Exploration of high-performing designs discovered in elastic and plastic materials. a**, Rendering of Willow, a high-performing design discovered using the plastic polymer polylactic acid (PLA) together with $\sigma$ vs. $\varepsilon$ for 15 identically prepared PLA Willow components. **b**, Rendering of Iroko, a high-performing design discovered using the hyperelastic polymer TPU-2, together with $\sigma$ vs. $\varepsilon$ for 15 identically prepared TPU-2 Iroko components. **c**, Measured $K_s^*$ vs. polymer elastic modulus $E$ for Iroko and Willow components made from one of five polymers. Dashed lines show a sigmoidal fit to guide the eye. Error bars represent one standard deviation.

While Willow and Iroko represent optimizations for the materials in which they were first discovered, the question remains of whether the performance of these shapes can translate to other materials or if it is a highly bespoke optimization of this combination of material and design. To explore this, components based on the Willow and Iroko designs were fabricated using a wide range of materials and tested in triplicate (Figure S13). Studying $K_s^*$ of these components showed the limitations of the transferability of these designs (Figure 3c). While each design performed with comparable $K_s^*$ for materials in their respective classes (*i.e.* hyperelastic vs. plastic), a transition region was observed at intermediate $E$. This observation reveals how material stiffness and plasticity modulate the energy-absorbing capacity of geometric designs. Specifically, higher stiffness together with greater plasticity mitigates the amount of softening the component exhibits as portions of it bend during compression. Overall, the comparison of Willow and Iroko confirmed that designs perform well within specific classes of materials, but that these geometric motifs do not directly translate to different material classes.

While the SDL-based campaign was able to discover highly efficient designs, we hypothesized that the broader corpus of mechanical tests performed during this campaign could provide further mechanical insight. As an initial exploration of this idea, the results of all the experiments performed with TPU-2 are shown in Figure 4a. The shaded region denotes the convex hull that estimates the space of accessible properties. This shows that the best performance observed for this material occurs at a single $\sigma_t^*$, which we denote $\sigma_{tp}$. Interestingly, all other materials studied exhibit a similarly shaped convex hull with a distinct peak (Figure S14), highlighting both the importance and the feasibility of tuning the material properties to the specific energy-absorbing application. We found that over the seven tested materials, $\sigma_{tp}$ was strongly correlated with the polymer plateau stress $\sigma_p$ (Figure 4b), providing an algorithmic process for selecting a material to optimally match use cases across a wide range of threshold stresses determined from different systems to be protected. Then the material can be selected and structured to maximize $K_s$ at that $\sigma_t$.

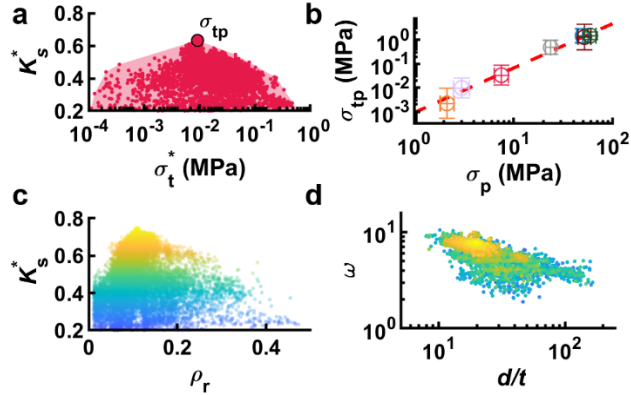

**Fig. 4 | Design insights that emerge from mechanical dataset. a**, Experimental data for TPU-2 with each dot representing $K_s^*$ vs. $\sigma_t^*$ for a specific experiment and the shaded region denoting the convex hull of the entire performance space corresponding to that material. The maximum point $\sigma_{tp}$ shows a critical stress $\sigma_t^*$ at which the highest $K_s^*$ is observed for that material. **b**, $\sigma_{tp}$ vs. polymer plateau stress $\sigma_p$ for seven polymers tested during the campaign together with a power law fit shown as a dashed line. Marker color indicates the material type. Error bars indicate one standard deviation of $\sigma_{tp}$ found throughout the campaign. **c**, $K_s^*$ vs. relative density $\rho_r$ for all components tested during the campaign with point color denoting $K_s^*$. **d**, Normalized height $\omega$ vs. $d/t$ for all components tested during the campaign in which point color denotes $K_s^*$ as in c.

In addition to tuning material properties, we hypothesized that unifying features of high-performing designs could be extracted to provide transferable guidance for realizing efficient structures. For example, it is reasonable to expect that the relative density $\rho_r$ of the component would strongly influence $K_s^*$. This hypothesis is motivated by the observation that the two factors that together bound $K_s^*$ are the flatness of the plateau region and the strain $\varepsilon_d$ where this plateau drastically rises due to densification. It has been observed for foams that $\varepsilon_d$ is bounded by relative density.[4,31] Because $\varepsilon_d \geq K_s^*$, we hypothesized that low $\rho_r$ is necessary for high $K_s^*$. Examining $K_s^*$ vs. $\rho_r$, we found that $K_s^*$ is peaked at $\rho_r \sim 0.1$ with all designs with $K_s^* \geq 65\%$ having $0.05 \leq \rho_r \leq 0.21$, providing guidance for structural design. Interestingly, because $\rho_r$ can be calculated prior to fabrication, limiting physical testing to designs with $\rho_r$ in this range can eliminate potential components that are unable to achieve high $K_s^*$.

Beyond the aggregate details of the design, there is a great deal of work exploring the mechanical regimes present for cylindrical shells under uniaxial compression. For example, the ratio $d/t$ of a cylindrical shell determines whether plastic cylinders fail through plastic deformation (thick wall limit) or fail elastically through the formation of local buckles (thin wall limit).[45] This transition has been observed to occur at $d/t \sim 100$. Further, the height of cylindrical shells is often characterized by the dimensionless length parameter $\omega = h/\sqrt{dt}$.[46] Here, cylinders are considered to be short when $\omega < 1.7$. Plotting $K_s^*$ vs. $d/t$ and $\omega$ reveals that all of the highest performing structures (*i.e.* $K_s^* \geq 70\%$), which were made from plastic materials, were concentrated in a narrow range around $d/t \sim 20$ and $\omega \sim 8$, which can be considered thick-walled medium-length cylindrical shells (Figure 4d). For simple cylinders in this region, one would expect elastic buckling and plastic deformation to play major roles. Thus, one way to understand the data-driven optimization process is that the other eight geometric parameters that define a GCS component have been tuned to guide these complex buckling and plastic interactions to interact constructively. Interestingly, we may use the tools of machine learning to identify which geometric motifs are most responsible for this improvement. In particular, we employ Shapley additive explanations (SHAP) analysis to find that the four-lobed profile of the cylinder ($x_{2-3}$) together with the linear and sinusoidal twist of this profile along the shell ($x_{6-8}$) are together responsible for 90% of the improvement over a simple cylindrical shell (Figure S15). Mechanically, this suggests that the key feature for improving the efficiency is producing local plastic deformation events that result in sufficient self-self contacts to prevent post-yield weakening. This feature allows the structure to maximize the material plasticity that occurs while maintaining a flat post-yield region.

Conclusion

This work reports a series of mechanical insights that resulted from performing an extensive experimental campaign using an SDL. Through the exploration of a vast parameter space, we were able to identify components with superlative $K_s$, advancing the frontier of energy absorption and finally overcoming the record previously held by nature. These designs were found to be general within material classes, showing the opportunities and limitations of transferability of the designs. This SDL campaign also illustrated how optimization and learning can be complementary goals in that the generated corpus of data allowed for the extraction of design insights for optimal use of polymers. These insights include matching polymer materials to target use cases, highlighting the use of $\rho_r$ as an aggregate descriptor, and gaining connections to the broader literature on mechanics of cylindrical shells. From a mechanics standpoint, we expect that this work will provide geometric motifs that lead to more efficient and safer protective equipment. From a broader learning perspective, this work shows how the iterative and collaborative interaction between SDLs and human teams can lead to sustained progress.

**Main References**

## Methods

### 1. Design of generalized cylindrical shells

To provide a vast space of potential designs that are topologically identical to cylindrical shells, we developed a generalized cylindrical shell (GCS) family of structures, which is an 11-dimensional parameter space (Figure S3). Three of these parameters are common to any cylindrical shell, namely the shell height $h$, wall thickness $t$, and average perimeter $P_0$. Beyond these three variables, eight additional parameters $\vec{x} = [x_1, x_2, x_3, x_4, x_5, x_6, x_7, x_8]$ are introduced that change the height-dependent cross section of the shell. The azimuthally-dependent radius $r(z,\phi)$ is shifted by adding two cosine functions with set periodicities as inspired by the summed cosine design of mechanical structures.[47] In particular, $r$ at any given height $z$ and azimuthal angle $\phi$ is given by

$$r(z, \phi) = r_0(z) \left[1 + C_4(z) \cos\left(4(\phi + \phi_0(z))\right) + C_8(z) \cos\left(8(\phi + \phi_0(z))\right)\right], \quad (1)$$

Where $C_4(z)$ and $C_8(z)$ are amplitude prefactors to the summed cosines, $\phi_0(z)$ is a rotational offset, and $r_0(z)$ is a prefactor adjusted to set the height-dependent perimeter $P(z)$ of the shell. Each of these functions is defined by terms of $\vec{x}$. Specifically, we define,

$$P(z) = P_0 \left[1 + x_1 \left(\frac{z}{h} - \frac{1}{2}\right)\right], \quad (2)$$

such that $x_1$ is the difference between the perimeter at the top of the component and the perimeter at the bottom. The Python function scipy.optimize.minimize was used to minimize the error between the $P(z)$ and the actual perimeter of Equation (1), estimated using Simpson's rule, by adjusting $r_0(z)$ at each layer.

Each summed cosine term is defined by specifying its value at the top and bottom of the shell and linearly interpolating between these points, specifically,

$$C_4(z) = x_2 \frac{h-z}{h} + x_3 \frac{z}{h}, \quad (3)$$

and

$$C_8(z) = x_4 \frac{h-z}{h} + x_5 \frac{z}{h}. \quad (4)$$

To determine the azimuthal offset of each layer, we include two ways in which this can vary with height, namely a linear and sinusoidal shift. Specifically, we define,

$$\phi_0(z) = x_6 \frac{z}{h} + x_7 \sin(2\pi x_8 z). \quad (5)$$

Code to generate standard triangle language (STL) models based on the GCS family of shapes is available https://github.com/bu-shapelab/gcs.

### 2. Sample Preparation

To fabricate samples, STL files were converted to G-code using Slic3r v.1.3.0. Filament rolls for 3D printing were purchased and used as received. They include three different types of thermoplastic polyurethane (TPU): TPU-1 (NinjaFlex-Ninjatek), TPU-2 (Cheetah-Ninjatek), and TPU-3 (Armadillo-Ninjatek). Additionally, four more filaments were used: thermoplastic elastomer (TPE) (Chinchilla-Ninjatek), nylon (Matterhackers Pro Series), polyethylene terephthalate glycol (PETG) (Matterhackers Pro Series), and polylactic acid (PLA) (eSun PLA+ and MakerGear). Samples were fabricated using MakerGear M3 printers with either a 0.5 or a 0.75 mm nozzle at 80 °C bed temperature, 250 °C nozzle temperature (except for PLA, which was printed using 220 °C), and 15 mm/s print speed using vase mode. The cylindrical shell sample in Figure 1A was fabricated out of TPU-2 (Cheetah – Ninjatek) to be 19.5 mm tall, 27.9 mm wide, and have 0.5 mm thick walls.

## 3. Mechanical Characterization of Samples

Quasi-static uniaxial compression was performed using an Instron 5965 with a 5 kN load cell at 2 mm/min until the force reached 4.5 kN or until the platens were separated by less than 0.4 mm. The resulting force-displacement data was converted to stress-strain by dividing the force by the area of a hexagon that would enclose the component and by dividing the displacement by the initial height, respectively (Figure S2). The mechanical energy absorbing efficiency $K_s$ vs. threshold stress $\sigma_t$ was computed by dividing the amount of energy absorbed below $\sigma_t$ by the maximum amount that could be absorbed without exceeding that stress.

To determine the mechanical properties of each roll of filament, solid cylinders (100% infill) were printed measuring 16 mm tall and 8 mm in diameter. These cylinders were then tested in uniaxial compression at 2 mm/min. Force-displacement curves were converted to stress-strain curves by dividing the force by the component cross-sectional area and by dividing the displacement by the height, respectively. From the resulting stress-strain curves, three material properties were calculated: the modulus of the material, plateau stress $\sigma_p$, and the rebound fraction. The modulus $E$ was calculated by fitting a series of lines in windows of 0.05 to 0.25 strain and an initial strain location of 0 to 0.25 strain (to avoid toe regions), both in increments of 0.05. The largest slope observed was taken as the modulus for the sample. The $\sigma_p$ was the stress value at 25% strain. The rebound fraction was the height of the part after a one-minute relaxation period divided by the initial height before testing, both measured by the Instron. One cylinder was tested for each roll of filament used.

## 4. Development of the Bayesian experimental autonomous researcher

In order to study the mechanical energy absorbed by additively manufactured components under uniaxial compression, we developed and utilized a self-driving lab (SDL). This system incorporated several distinct instruments, computers, and algorithms that worked in concert to select experiments, construct physical samples, and test them with minimal human intervention. From a hardware perspective, this system consisted of five fused filament fabrication (FFF) printers (MakerGear M3-ID) arrayed in an arc. In the center of this arc was a 6-axis robot arm (Universal Robotics UR5e). Also in the working radius of this arm was a scale (Sartorius CP225D) and a universal testing machine (UTM) (Instron 5965). The arm had a webcam (Logitech C930e) to allow for monitoring the flow of experiments and there was a video camera (PixelLINK PL-D722) with lens (Infinity InfiniMite Alpha) mounted facing the bottom platen of the UTM to record videos of the tests. A picture of the SDL is shown as Figure S5. The hardware and software organization of this system is shown in Figure S7.

A flow chart describing the core actions of this system is shown in Figure S8. At the most basic level, the system comprised a loop implemented in MATLAB 2021a (Mathworks Inc) in which the system repeatedly iterated through six potential actions, namely: (1) Use the accumulated data to select the design and material to be tested next given the available printer and materials. (2) Generate the digital design files needed to run the available printer. (3) Send the G-code file to the printer and begin printing the component. (4) Retrieve the completed component from the printer and weigh it using the scale. (5) Retrieve the component from the scale, place it on the platen of the universal testing machine, run the mechanical testing program, and then clear the component from the platen. (6) Read the results of the mechanical testing and update the accumulated data. The order of priority was tuned to maximize the throughput of the system by giving priority to actions that were likely to become bottlenecks. The details of these steps are given in the supporting information.

## 5. The research campaign

Over the course of the campaign, 25,387 experiments were performed (Figure S9). Although individual experiments were rarely selected by hand, a variety of decisions were made by the researchers along the way (Figure S10). Changes were made to the parameter space under consideration, such as adding sinusoidal twist or switching to Latin hypercube sampling (LHS) (Figure S10a). Additionally, new materials were added to the campaign, and the mix of filaments loaded into the printers was adjusted to focus on specific goals (Figure S10b). Finally, various decision policies were used, including maximum variance, expected improvement, and upper confidence bound (Figure S10c).

## Acknowledgements

This work was supported by Google LLC, the Boston University Rafik B. Hariri Institute for Computing and Computational Science & Engineering (2017-10-005), the National Science Foundation (CMMI-1661412), and the US Army CCDC Soldier Center (contract W911QY2020002). This work was funded by Honeywell Federal Manufacturing and Technologies through contract number N000471618. Honeywell Federal Manufacturing and Technologies, LLC operates the Kansas City National Security Campus for the United States Department of Energy/National Nuclear Security Administration under contract number DE-NA0002839

## Author Contributions

**Kelsey L. Snapp:** Conceptualization, Methodology, Software, Investigation, Writing-Original Draft. **Benjamin Verdier**: Conceptualization, Methodology, Software, Writing-Review & Editing. **Aldair Gongora:** Conceptualization, Methodology, Software, Writing-Review &

Editing. **Samuel Silverman:** Software, Visualization, Writing-Review & Editing. **Adedire D. Adesiji:** Investigation, Writing-Review & Editing. **Elise F. Morgan:** Formal Analysis, Writing-Review & Editing. **Timothy J. Lawton:** Conceptualization, Methodology, Writing-Review & Editing. **Emily Whiting:** Conceptualization, Methodology, Writing-Review & Editing, Supervision. **Keith A. Brown:** Conceptualization, Methodology, Writing-Original Draft, Writing-Review & Editing, Supervision

**Competing Interest Declaration**

The authors declare no competing interests.

**Additional Information**

Correspondence and requests for materials should be addressed to Keith Brown.

The raw data associated with these experiments is shared via kablab.org/data.

**Supplementary Information**

**Autonomous Discovery of Tough Structures**

Kelsey L. Snapp[1], Benjamin Verdier[2], Aldair Gongora[1], Samuel Silverman[2], Adedire D. Adesiji[1], Elise F. Morgan[1,3,4], Timothy J. Lawton[5], Emily Whiting[2], and Keith A. Brown[1,3,6]

1 Department of Mechanical Engineering, Boston University, Boston, MA, USA

2 Department of Computer Science, Boston University, Boston, MA, USA

3 Division of Materials Science & Engineering, Boston University, Boston, MA, USA

4 Department of Biomedical Engineering, Boston University, Boston, MA, USA

5 Soldier Protection Directorate, US Army Combat Capabilities Development Command Soldier Center, Natick, MA, USA

6 Physics Department, Boston University, Boston, MA, USA

**Table of Contents**

# 1. Definition and Records of Mechanical Energy Absorbing Efficiency

The energy absorbing efficiency $K_s$ of a material or structure in compression can be calculated by dividing the amount of energy absorbed before surpassing a stress threshold $\sigma_t$ by the maximum energy that could be absorbed below that threshold, i.e. compressing to an engineering strain of 1 while maintaining an engineering stress $\sigma = \sigma_t$ (Figure 1a). Equivalently, $K_s$ can also be calculated directly from force-displacement data by calculating the amount of energy absorbed before surpassing a force threshold by the maximum amount of energy that could be absorbed below that threshold, i.e. compressing until the displacement of the component equals its initial height while maintaining a force equal to the threshold force. For most structures, $K_s$ reaches its maximum energy absorbing efficiency $K_s^*$ at a single ideal threshold stress $\sigma_t^*$. This single point, ($\sigma_t^*$, $K_s^*$) can be used to describe the ideal operating performance of a structure or material. To illustrate common and superlative values of $K_s^*$, a summary of literature values is shown graphically in Figure S1. The sources of these points are given in Table S1. Values of $\sigma_t^*$ and $K_s^*$ not directly reported were computed based on reported force-displacement or stress-strain plots. Included on this plot are two values taken from this work.

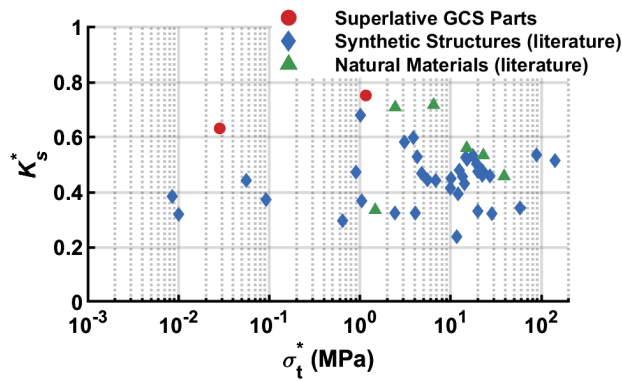

**Fig. S1 | Common and superlative structures and materials.** Synthetic structures (blue diamond) and natural materials (green triangle) gathered from literature, with superlative plastic and hyperelastic generalized cylindrical shells (GCS-this study) components (red circle).

| $\sigma_t^*$ (MPa) | $K_s^*$ (%) | Material origin | Reference |
|---|---|---|---|
| 8.50×10⁻³ | 38.5 | Synthetic | https://doi.org/10.1016/j.matdes.2017.11.037 |
| 1.00×10⁻² | 32.0 | Synthetic | https://doi.org/10.1016/j.matdes.2017.11.037 |
| 2.85×10⁻² | 63.2 | Synthetic | This work – ADTS ID 22335 |
| 5.55×10⁻² | 44.4 | Synthetic | https://doi.org/10.1016/j.matdes.2017.11.037 |
| 9.17×10⁻² | 37.4 | Synthetic | https://doi.org/10.1016/j.matdes.2017.11.037 |
| 6.47×10⁻¹ | 29.7 | Synthetic | https://doi.org/10.1002/admt.201800419 |
| 9.01×10⁻¹ | 47.3 | Synthetic | https://doi.org/10.1002/admt.201800419 |
| 1.01 | 68.1 | Synthetic | https://doi.org/10.1177/0021955X06063519 |
| 1.04 | 36.9 | Synthetic | https://doi.org/10.1016/j.actamat.2004.05.039 |
| 1.16 | 75.2 | Synthetic | This work – ADTS ID 21285 |
| 1.47 | 33.5 | Natural | https://doi.org/10.1016/j.jmbbm.2019.103603 |
| 2.42 | 32.3 | Synthetic | https://doi.org/10.1177/0021955X06063519 |
| 2.45 | 70.9 | Natural | https://doi.org/10.1016/S0167-6636(02)00268-5 |
| 3.08 | 58.2 | Synthetic | https://doi.org/10.1016/j.msea.2004.03.051 |
| 3.89 | 59.8 | Synthetic | https://doi.org/10.1177/0021955X06063519 |
| 4.07 | 32.4 | Synthetic | https://doi.org/10.1177/0021955X06063519 |
| 4.30 | 52.9 | Synthetic | https://doi.org/10.1016/j.msea.2004.03.051 |
| 4.76 | 46.8 | Synthetic | https://doi.org/10.1016/j.actamat.2004.05.039 |
| 5.53 | 44.5 | Synthetic | https://doi.org/10.1016/j.ijimpeng.2010.03.007 |
| 6.47 | 71.8 | Natural | https://doi.org/10.1016/S0167-6636(02)00268-5 |
| 6.88 | 44.2 | Synthetic | https://doi.org/10.1177/0021955X06063519 |
| 9.99 | 41.5 | Synthetic | https://doi.org/10.2140/jomms.2013.8.65 |
| $1.01 \times 10^1$ | 45.0 | Synthetic | https://doi.org/10.1016/j.actamat.2004.05.039 |
| $1.17 \times 10^1$ | 23.8 | Synthetic | https://doi.org/10.1002/admt.201800419 |
| $1.21 \times 10^1$ | 39.4 | Synthetic | https://doi.org/10.1177/0731684419868018 |
| $1.24 \times 10^1$ | 47.9 | Synthetic | https://doi.org/10.1177/0731684419868018 |
| $1.34 \times 10^1$ | 45.6 | Synthetic | https://doi.org/10.1177/0731684419868018 |
| $1.42 \times 10^1$ | 43.2 | Synthetic | https://doi.org/10.1177/0731684419868018 |
| $1.46 \times 10^1$ | 52.8 | Synthetic | https://doi.org/10.1177/0731684419868018 |
| $1.50 \times 10^1$ | 56.1 | Natural | https://doi.org/10.1016/S0167-6636(02)00268-5 |
| $1.52 \times 10^1$ | 51.9 | Synthetic | https://doi.org/10.1177/0731684419868018 |
| $1.74 \times 10^1$ | 53.3 | Synthetic | https://doi.org/10.1177/0731684419868018 |
| $1.92 \times 10^1$ | 50.3 | Synthetic | https://doi.org/10.1177/0731684419868018 |
| $1.98 \times 10^1$ | 47.6 | Synthetic | https://doi.org/10.1177/0731684419868018 |
| $1.98 \times 10^1$ | 33.2 | Synthetic | https://doi.org/10.1002/admt.201800419 |
| $2.22 \times 10^1$ | 47.7 | Synthetic | https://doi.org/10.2140/jomms.2013.8.65 |
| $2.22 \times 10^1$ | 46.1 | Synthetic | https://doi.org/10.1177/0731684419868018 |
| $2.25 \times 10^1$ | 47.5 | Synthetic | https://doi.org/10.2140/jomms.2013.8.65 |
| $2.29 \times 10^1$ | 53.5 | Natural | https://doi.org/10.1016/S0167-6636(02)00268-5 |
| $2.69 \times 10^1$ | 45.8 | Synthetic | https://doi.org/10.2140/jomms.2013.8.65 |
| $2.86 \times 10^1$ | 32.1 | Synthetic | https://doi.org/10.1177/0021955X06063519 |
| $3.88 \times 10^1$ | 45.8 | Natural | https://doi.org/10.1016/S0167-6636(02)00268-5 |
| $5.84 \times 10^1$ | 34.3 | Synthetic | https://doi.org/10.1016/j.ijsolstr.2015.02.020 |
| $8.86 \times 10^1$ | 53.6 | Synthetic | https://doi.org/10.1016/j.msea.2004.03.051 |
| $1.40 \times 10^2$ | 51.5 | Synthetic | https://doi.org/10.1016/j.msea.2004.03.051 |

**Table S1 | Common and superlative structures and materials**

## 2. Process for Converting Force-Displacement into Stress-Strain

When converting from force-displacement curves to engineering stress-engineering strain curves (simply called stress and strain hereafter), it is necessary to define the area of the component and its height. For traditional materials, this process is straightforward as it amounts to defining the cross-sectional area of the component under study. However, for more complex structures, the area of the component is less clear. Here, we define the cross-sectional area to be the amount of area that is required per component to tile the component infinitely on a plane. To calculate this algorithmically, we used the following steps (illustrated graphically in Figure S2):

1. Find the maximum radius $r_{max}$ of the component by finding the maximum of the radius $r$ at all heights $z$ and azimuthal angles $\phi$, as defined in Equation (1) in the Methods.
2. Enclose the component with a cylinder with radius $r_{max}$ (Figure S2b).
3. Enclose the component in a hexagonal prism that circumscribes the cylinder (Figure S2c).
4. The area of the hexagonal prism is used as an estimate of the cross-sectional area needed to tile the component on a plane.

Although some designs may be more closely packed in a square lattice, most are more closely packed using the hexagonal approach (Figure S2d) due to the applied linear and sinusoidal twists, and this approach is invariant of the rotational orientation of the design. Therefore, this hexagonal packing approach was used to estimate the cross-sectional area of all designs.

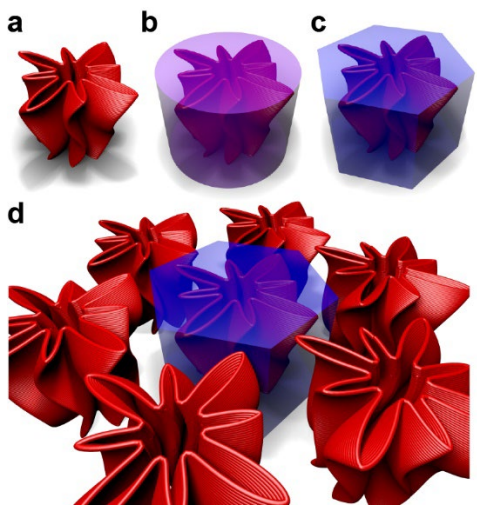

**Fig. S2 | Calculating the effective area of GCS designs. a-b,** To calculate the effective cross-sectional area of a design, it is fit into a cylinder based on its maximum radius. **c,** This cylinder is then enclosed in a hexagonal prism. **d,** The hexagonal prism can be tiled infinitely in a plane. Thus, the effective cross-sectional area of a design is estimated as the area necessary per design to tile it in a plane without collisions.

### 3. Defining a Generalized Cylindrical Shell

A cylindrical shell is often defined in terms of its height $h$, wall thickness $t$, and diameter $d$. Here, we design generalized cylindrical shells (GSC) that are topologically consistent with cylindrical shells and have a consistent wall thickness and height, but vary in their cross-sectional profile along the axial direction. As a diameter is not an appropriate measure for such a complex shape, we parameterize these using their average perimeter $P_0$. A GCS design is realized by deforming cylindrical shells using three distinct transformations: variable perimeter, variable cross section, and twist (Figure S3a). These transformations are defined mathematically in the methods section of the main text. Briefly, the variable perimeter is realized by linearly varying the perimeter from the top of the GCS to the bottom of the GCS (Figure S3b). In this campaign, the perimeter of the top was constrained to be larger than the perimeter of the bottom to ease with component removal. The cross sections of the GCS were transformed using a summed cosine function (Figure S3c). The top cross section and bottom cross section are specified, and each intermediate layer is calculated as a linear interpolation of these two faces, ensuring a manifold surface. Finally, both sinusoidal and linear twist can be applied to these cross sections in a height-dependent manner (Figure S3d). Collectively, these transforms allow for more than trillions of unique designs.

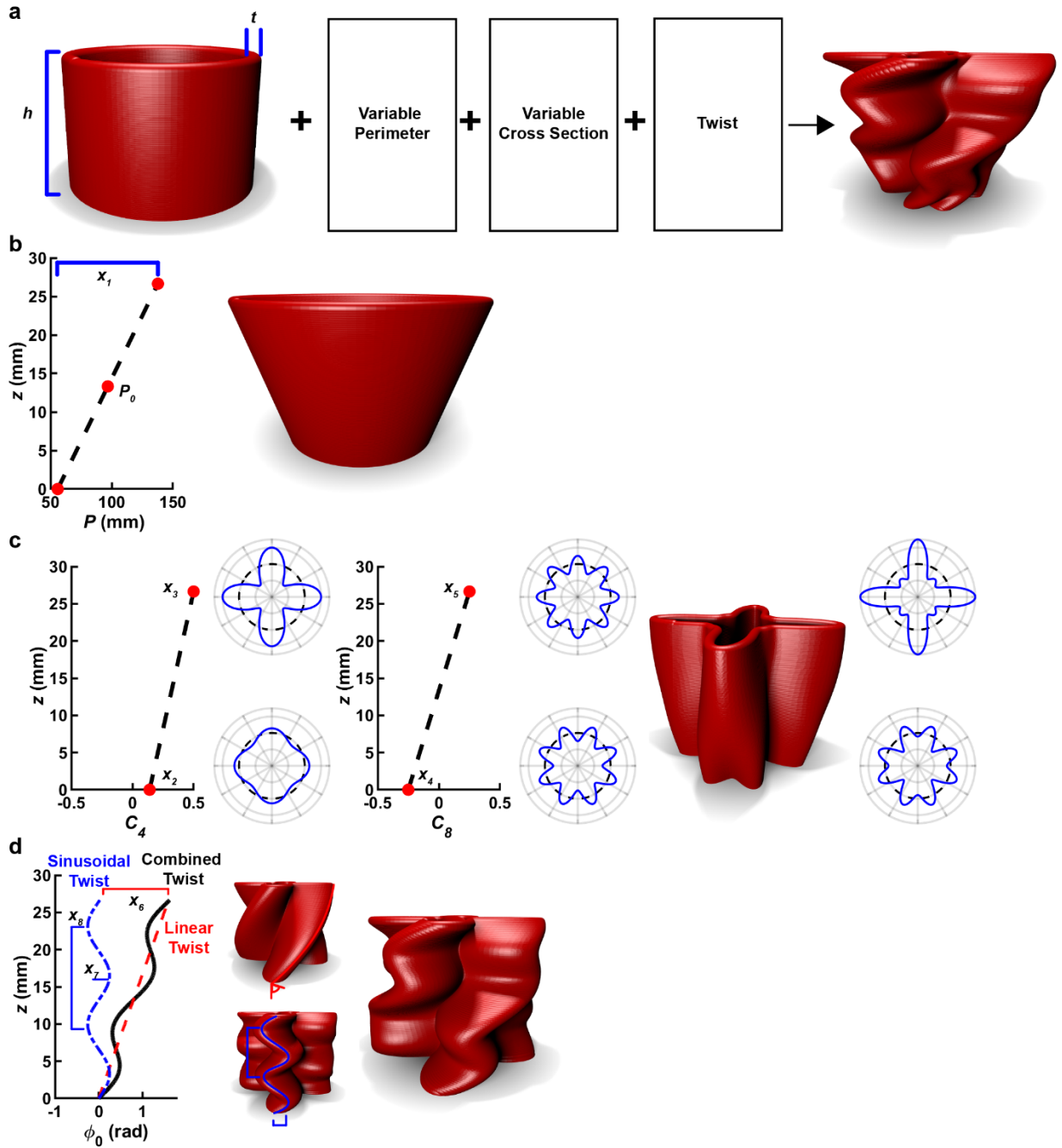

**Fig. S3 | Generalized cylindrical shells. a,** Generalized cylindrical shells (GCS) are realized by transforming a cylindrical shell of height $h$ and wall thickness $t$ to create interesting shapes that preserve the topology of the shell. **b,** The perimeter $P$ varies linearly along the height $z$ of the shell based on an average perimeter $P_0$ and a perimeter difference $x_1$. **c,** The cross sections of each layer are deformed in a $z$-dependent manner using a summed cosine function with 4-period amplitude $C_4$ and 8-period amplitude $C_8$. These are defined at the top and bottom by four variables $x_2$, $x_3$, $x_4$, and $x_5$, and linearly interpolated to determine the cross section at any $z$. **d,** The cross sections of the design are rotated about the cylinder axis in a $z$-dependent manner by rotation angle $\phi_0$ using both linear and sinusoidal twists as defined by linear twist $x_6$, sinusoidal twist amplitude $x_7$, and sinusoidal twist period $x_8$.

## 4. Polymers under Consideration

The polymer materials studied in this work are provided in Table S2 along with the temperature at which they were printed, the temperature at which the print bed was held during removal, and the material class. In addition, for each spool of material studied, a cylindrical sample was printed and tested to estimate the material properties of the polymer. The details of this process are provided in the methods. As shown in Figure S4, the result of this testing are estimates of the elastic modulus $E$ and plateau stress $\sigma_p$ of each material. In addition, the degree to which the cylinder rebounded after a one-minute relaxation period was also recorded, although this is an imprecise measure of elasticity as a consistent force threshold was used for all tests, indicating that different materials experienced different total strains. Nevertheless, the plastic materials rebounded less than the hyperelastic materials, despite their total strain being lower.

| Material | Manufacturer | Nozzle Temperature (°C) | Bed Removal Temperature (°C) | Class | Spools Used |
|---|---|---|---|---|---|
| TPE (Chinchilla) | NinjaTek | 250 | 100 | Hyperelastic | 9 |
| TPU-1 (NinjaFlex) | NinjaTek | 250 | 100 | Hyperelastic | 16 |
| TPU-2 (Cheetah) | NinjaTek | 250 | 100 | Hyperelastic | 37 |
| TPU-3 (Armadillo) | NinjaTek | 250 | 30 | Intermediate | 11 |
| Nylon | MatterHackers | 250 | 30 | Plastic | 2 |
| PETG | MatterHackers | 250 | 30 | Plastic | 5 |
| PLA | eSun/MakerGear | 220 | 30 | Plastic | 29 |

**Table. S2 | Filaments studied in this work along with their processing settings.**

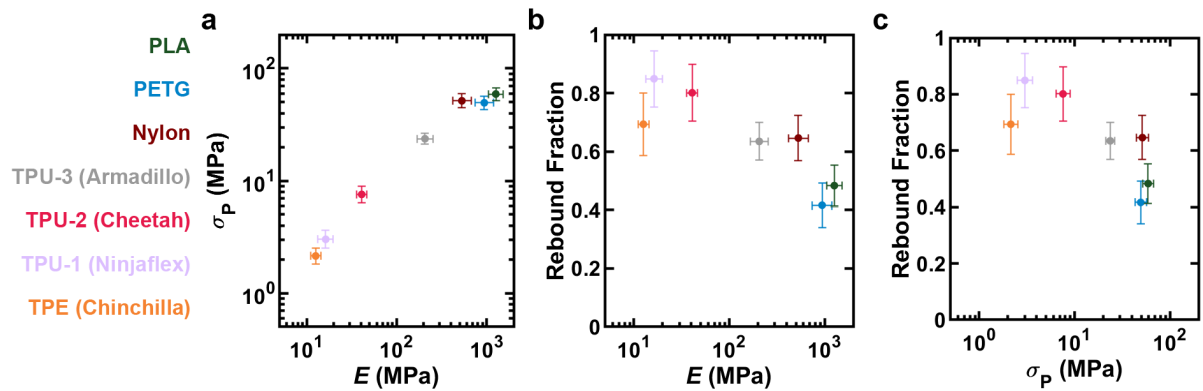

**Fig. S4 | Material characterization of polymers studied. a,** Plateau stress $\sigma_p$ vs. elastic modulus $E$ for seven materials used in this campaign. **b,** Rebound fraction vs. $E$. **c,** Rebound fraction vs. $\sigma_p$. Error bars represent one standard deviation. Here, $\sigma_p$ is the stress at 25% strain. Rebound fraction is the height after 1 minute relaxation divided by the initial height.

## 5. The Bayesian Experimental Autonomous Researcher

The Bayesian experimental autonomous researcher (BEAR) consists of a collection of computers and other hardware that work together to perform research experiments without direct human intervention. It consists of five fused filament fabrication 3D printers, a scale, a universal testing machine, and a six-axis arm to transfer experiments between the different stations (Figure S5). The various components are controlled centrally by a custom-made MATLAB script (Figure S7). The BEAR has a series of tasks that it can do, which it does in order of a user-modifiable priority (Figure S8).

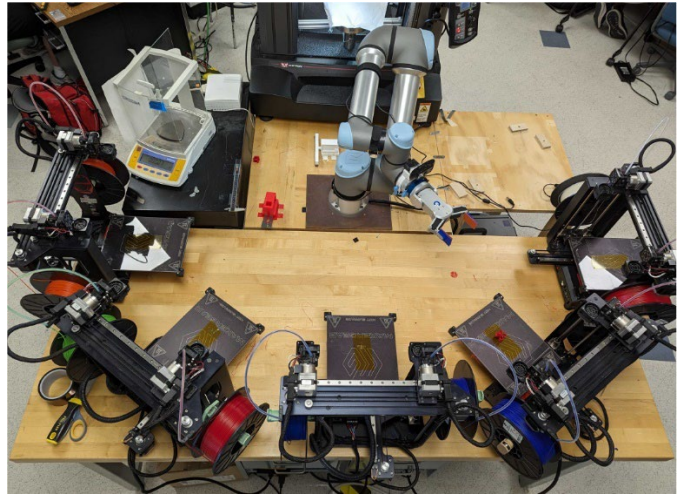

**Fig. S5 | Picture of the Bayesian experimental autonomous researcher (BEAR), consisting of five fused filament fabrication 3D printers, a six-axis robot arm, a scale, and a universal testing machine.**

### 5.1 Select Experiment

Bayesian optimization was used by the BEAR to algorithmically select additional experiments. This process includes the conditioning of a surrogate model to approximate the connection between input space and output space and then the use of an acquisition function to evaluate this model to find experiments that are believed to be most useful to perform. Since the goal of this work was to identify structures with high $K_s^*$, we treated this as a maximization problem. The input space for this maximization was both the design of the GCS and the material used to realize a component out of this design. As such, we required a 13-dimensional input (11 geometric parameters and two material properties). For the output space, we were not just interested in $K_s^*$, but we also found it necessary to predict both $\sigma_t^*$ and whether the component (design plus material) could be fabricated. Gaussian process regressions (GPRs) were used to predict $K_s^*$ and $\sigma_t^*$. A neural network with one hidden layer equal to the input size was used to predict component printability.

In developing surrogate models, transformations were done to the GCS design parameters and material parameters to improve the accuracy of the models. The overall motivation of these transformations was to improve correlations in the input space and thus improve predictions in the output space. The full list of the transformed input and output spaces are given in Tables S3 and S4. For example, the logarithms of $\sigma_p$, $\sigma_t^*$, and $E$ were taken because their values varied over several orders of magnitude and the points were more evenly spaced when considered logarithmically rather than linearly. Additionally, rather than specify $P_0$, we preferred to specify the target mass $m$ normalized by $h$, or the mass per height $m/h$. Additionally, a wall angle $\theta$ was used as it was hypothesized that the angle of the wall was more important than the absolute value of the change in $P$. This wall angle was estimated using the formula $\theta = \mathrm{atan}\left(\frac{x_1}{2\pi h}\right)$. Finally, rather than conditioning the GPR to directly predict $K_s^*$, we found that it was useful to transform $K_s^*$ to emphasize differences at the high end while minimizing differences at the low end and to explicitly prevent the model from predicting physically impossible values (i.e. $K_s^*>1$ or $K_s^*<0$). Thus, we instead predicted $\mathrm{atanh}(2K_s^*-1)$. This function was chosen because it monotonically transforms inputs from 0-1 to outputs from negative infinity to positive infinity.

| Model input variable | Description |
| --- | --- |
| $h$ | Height |
| $m/h$ | Mass per height |
| $t$ | Wall thickness |
| $\mathrm{atan}\left(\frac{x_1}{2\pi h}\right)$ | Wall angle |
| $x_2$ | 4-period amplitude of bottom cross section |
| $x_3$ | 4-period amplitude of top cross section |
| $x_4$ | 8-period amplitude of bottom cross section |
| $x_5$ | 8-period amplitude of top cross section |
| $x_6/h$ | Linear rotation per height |
| $x_7$ | Sinusoidal rotation amplitude |
| $x_8$ | Sinusoidal rotation wavelength |
| $\ln(E)$ | Natural log of the polymer elastic modulus |
| $\ln(\sigma_p)$ | Natural log of the polymer plateau stress |

**Table. S3 | Inputs to the machine learning models used for Bayesian optimization.**

| Model output variable | Description | Model type |
| --- | --- | --- |
| $\mathrm{atanh}(2K_s^*-1)$ | Transformed peak energy absorbing efficiency | Gaussian process regression |
| $\log\left(\frac{\sigma_t^*}{E}\right)^{0.408}$ | Log 10 of the ideal threshold stress normalized by the modulus and raised to an empirically determined power | Gaussian process regression |
| $p$ | Printability | Artificial neural network |

**Table. S4 | Outputs of the machine learning models used for most of the experimental campaign.**

When selecting a subsequent experiment for a given printer, not all combinations of designs and materials were available. Specifically, each printer had two independent extruders, which allowed two different filaments to be loaded at once. Further, each extruder had either a 0.5 mm diameter nozzle or a 0.75 mm diameter nozzle. Values of $t$ different from these diameters could be achieved by over or under extruding. We restricted $t < 0.7$ mm for the 0.5 mm diameter nozzle and $t \geq 0.7$ mm for the 0.75 mm nozzle.

To select an experiment, we define an acquisition function $a$ that takes as its input positions in parameter space along with the current surrogate models and select the experiment that maximizes $a$. Throughout the campaign, three types of acquisition functions were used: maximum variance ($a$ is equal to the variance in predicting $K_s^*$), expected improvement ($a$ is the predicted amount of improvement beyond the previous best $K_s^*$), and upper confidence bound ($a$ is the weighted sum of the predicted of $K_s^*$ and the predicted uncertainty in predicting $K_s^*$). The combination of $a$ and the strategy for finding its maximum is considered a decision policy. However, this process was not treated as a simple single-objective maximization. For instance, in all cases, $a$ is multiplied by the predicted printability $p$ to ensure that we are only considering components that are expected to be realizable in practice. Additionally, many of the decision policies are multi-objective, trying to find high values of $K_s^*$ across a range in $\sigma_t^*$. When this was the case, multiple GPR model predictions were combined to select a component by penalizing the $K_s^*$ prediction by the distance of its predicted $\sigma_t^*$ from the target $\sigma_t$ or by comparing the predicted $K_s^*$ to the performance of other tests at that $\sigma_t^*$. A full list of considered decision policies is given in Table S5. These policies were added sequentially during the progression of the campaign, so their order reflects the evolution of our thought process during the campaign, discussed further in Section 6. Additionally, the GPR models can be retrained using only data from the region of interest, which was begun with decision policy 19. This allowed the GPR to capture finer correlations in the parameter space around the region of interest. All models were trained using MATLAB's built in functions and the code is available at https://github.com/KelseyEng/BEAR_ADTS. Model Training was performed on Boston University's Shared Computing Cluster, where multiple compute nodes could work in parallel. GPR processing time scales with the number of experiments cubed.[1] Therefore, the longer the campaign ran, the more computationally expensive model building and component selection became.

| Decision Policy Number | Acquisition function | Metric | Number of Valid Experiments |
|---|---|---|---|
| 0 | Manually Selected | Researcher intuition or performance validation | 730 |
| 1 | Upper confidence bound | Full integral of force-displacement curve | 24 |
| 2 | Maximum variance | Full integral of force-displacement curve normalized by component mass | 916 |
| 3 | Expected improvement | Full integral of force-displacement curve normalized by component mass | 775 |
| 4 | Expected improvement | Expected acceleration of a simulated impact test | 93 |
| 5 | Expected improvement | $K_s$ at a target $\sigma_t$ | 249 |
| 6 | Expected improvement | $K_s^*$ penalized by an amount proportional to the distance between $\sigma_t^*$ and a target $\sigma_t$ | 97 |
| 7 | Expected improvement | $K_s^*$ penalized by an amount proportional to the distance between $\sigma_t^*$ and a target $\sigma_t$ with uncertainty in $\sigma_t^*$ considered | 383 |
| 8 | Expected improvement | $K_s^*$ minus the best $K_s$ previously observed at the predicted $\sigma_t^*$ | 3,219 |
| 9 | Expected improvement | $K_s^*$ minus the best $K_s$ previously observed at the predicted $\sigma_t^*$, but with limits imposed on the largest and smallest stresses considered | 31 |
| 10 | Expected improvement | $K_s^*$ minus the best $K_s$ previously observed at the predicted $\sigma_t^*$, but only considered components that could have been printed using the specific printer under consideration | 501 |
| 11 | Maximum variance | $K_s^*$, but only considering cylindrical shells | 34 |
| 12 | Expected improvement | $K_s^*$ | 1,608 |
| 13 | Not Used | | |
| 14 | Expected improvement | A weighted sum of the acceleration from a simulated impact test and the plateau stress of the component | 41 |
| 15 | Expected improvement | A weighted sum of the acceleration from a simulated impact test and the plateau stress of the component (different simulation model from DP 14) | 212 |
| 16 | Expected improvement | $K_s^*$ times the ideal threshold force for that component | 22 |
| 17 | Expected improvement | $K_s^*$ minus the best $K_s^*$ that could have been printed using the specific printer under consideration | 1,041 |
| 18 | Expected improvement | $K_s^*$, but only considering components near the best previously found component | 1,569 |
| 19 | Upper confidence bound | $K_s^*$, but only considering components near the best previously found component | 224 |
| 20 | Upper confidence bound | $K_s^*$, but only considering components near the best previously found component that have effective densities $\rho_d$ below 10% | 542 |
| 21 | Not Used | | |
| 22 | Expected improvement | $K_s^*$ minus the best $K_s$ previously observed at the predicted $\sigma_t^*$, but only considering components that could be continuously extruded without a linear twist | 523 |
| 23 | Expected improvement | $K_s^*$ minus the best $K_s$ previously observed at the predicted $\sigma_t^*$, but only considering components that could be continuously extruded with a linear twist | 286 |
| 24 | Expected improvement | $K_s^*$ of a two-component system | 129 |
| 25 | Expected improvement | $K_s^*$ of a two-component system, but only considering components near the best previously found pair of components | 1 |

**Table. S5 | Descriptions of decision policies used during campaign.**

Initially, sampling points were selected on a grid. Starting at ID 9,261, potential sampling points were selected using Latin hypercube sampling (LHS) to facilitate exploring space more finely. Starting with ID 11,763, after the proposed experiment had been selected, a second round of sampling points were added that were zoomed in a hypercube around the selected point to more closely find the maximum of *a*.

## 5.2. Generate G-code

Once a component has been selected for testing, the STL was generated using a custom Python script. This Python script (Python version 3.8.3) was run on the main computer and called from MATLAB using the command line function. The resulting STL was created as a solid object. In order to convert this STL file into the G-code needed for the printer, Slic3r (version 1.3.0) was run from the command line of MATLAB. Prior to sending the STL file, the Slic3er configuration file was edited using string manipulation directly from MATLAB to set the nozzle temperature, bed removal temperature, and extrusion multiplier. The two temperatures were designated by the human team based on our experience with these materials (see Table S2) while the extrusion multiplier was set as part of a feedback system to maintain component weight (see below). Slic3er was configured to use vase mode (spiral mode), which removes the tops and bottoms of solid objects and turns the STL solid into a shell. The output of this process is G-code for the print and predicted amount of filament that is needed to print this component, which is read into MATLAB. Using an initial set of calibration prints and subsequent use of integral feedback, we predicted the mass of the component from the amount of filament predicted to be used by the slicer. Adjusting the extrusion multiplier and reslicing the component provided a reliable method of controlling the mass of the final component (Figure S6). This gives effective control over $t$ by over or under expanding the material leaving the nozzle. It also allows the computer to automatically compensate for variations in the thickness of the filament diameter or variations between the stepper motor of different printers. Relatively slow print speeds of 15 mm/min were employed to prevent clogging, which was especially important for the softer filaments.

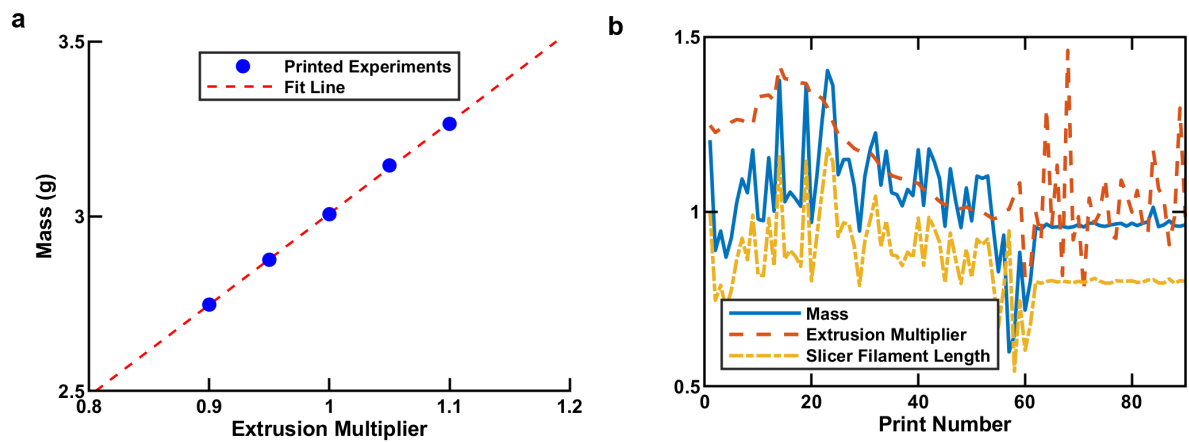

**Fig. S6 | Mass calibration through extrusion multiplier adjustments. a,** A single component printed five times with different extrusion multiplier shows that component mass is linearly correlated with extrusion multiplier. **b,** Applying integral tuning to the extrusion multiplier was ineffective because of variations in slicing complex curved structures, as seen in print number < 60. However, when integral tuning was applied to the slicer filament length by adjusting the extrusion multiplier rapidly, consistent mass was obtained (print number > 60). Mass is normalized by target mass and slicer filament length is normalized by the initial slicer filament length (print number = 1).

### 5.3 Begin Experiment

Once the G-code had been prepared for a given printer, the physical experiment was ready to begin. First, the arm moved into position over the chosen printer and the arm-mounted camera took a photograph of the print bed to ensure that it was free from debris and ready for the next print. To accomplish this, the picture was run through a neural net based on GoogLeNet[2] and classified as 'clear' or 'needs cleaning'. If the bed needed cleaning, the robot arm picked up a scraper and scraped the print bed. A new picture was then taken to verify that the print bed was clean. If the bed was still unclean, the BEAR would attempt to clean the bed up to ten times with the scraper. If the bed was still unclean, the printer would be deactivated and the operator was notified to clean the bed before further experiments. Once the bed was determined to be clear, the system sent G-code to the printer using OctoPrint.

### 5.4 Weigh Component

When a print was complete, as determined by querying the state of the printer through OctoPrint, the bed was heated (TPE, TPU-1-2) or allowed to cool (PLA, PETG, and Nylon, TPU-3) to facilitate removal of the component.[3] Once the desired temperature had been reached, the robot arm removed the component from the printer and moved it to the scale, which determined its mass. This mass reading was read through a serial port by MATLAB. If no mass was registered on the scale, the system attempted to re-grab the component from the print bed up to three times. At this stage, the arm-mounted camera took a photograph of the component on the scale, which was used to verify that the component was fully on the scale. Components that were misoriented, as determined by machine vision, were discarded before testing.

### 5.5 Test Component

If a component was on the scale and ready to be tested while the universal testing machine (UTM) was not performing any experiments, the component was moved to the UTM for compression testing. Once the component was in position, the main computer sent a command to the Instron computer to begin the test through a .mat file transferred by the cloud. The Video computer then told the Instron to start the test while it recorded a video of the compression testing. The test began with the top platen ~200 mm over the component. After zeroing the force sensor, the top platen moved at a rate of 50 mm/min toward the component until the force sensor registered 1 N. The platen then moved away from the component 1 mm so that it no longer was in contact with the component. At this point, the UTM started recording the force measurement while it lowered the top platen at 2 mm/min. A given test ended when either 1) the force exceeded the 4.5 kN force limit or 2) the top platen position fell below the safe height of 0.4 mm separation between the two platens. After a one minute relaxation period, the platen was lowered again at a rate of 100 mm/min until the force exceeded 1 N to find the rebound height. After testing, the component was removed from the UTM and stored. The platen was then cleaned with the robot arm to ensure that the platens were clear and ready for the next test. Each mechanical test took approximately ten minutes. A third computer recorded the Instron data and saved it to the cloud. When the test was finished, the Video computer informed the main computer that the UTM was now free for another experiment.

## 5.6 Process Results

When new experimental results were available to be processed, the raw force-displacement data was loaded into MATLAB. The as-printed height of the component was calculated by finding the platen separation when the moving median of twenty force measurements surpassed 0.3 N. The effective area of the component was calculated by computing the maximum radius of any layer of the component and using that as the apothem (distance from center to midpoint of side) of a circumscribing hexagon (Figure S2). Using this height and effective area, the force-displacement curve was converted to a stress $\sigma$ – strain $\varepsilon$ curve.

From the $\sigma$–$\varepsilon$ curve, a variety of useful metrics were calculated. To find $K_s^*$ and $\sigma_t^*$, $K_s$ was calculated at 1,400 $\sigma_t$ values that were logarithmically spaced between 10 Pa and 100 MPa. Ten additional $\sigma_t$ sampling points were selected by diving the $\sigma$–$\varepsilon$ curve into ten equal sections in $\varepsilon$ and finding the maximum $\sigma$ in each section. Because $\sigma_t^*$ is often a peak early in the $\sigma$–$\varepsilon$ curve, these ten additional sampling points can often determine $\sigma_t^*$ precisely. The densification strain $\varepsilon_d$ is the $\varepsilon$ value at which $\sigma$ first exceeds $\sigma_t^*$. The relative density of the component $\rho_r$ was calculated by dividing the mass of the component by the mass of solid material equal to the volume of the enclosing hexagon (Figure S2).

Finally, quality control checks were performed to determine if the sample should be included in the complete dataset. Components that were not within 5% of their mass target or within 5% of their target height were excluded from the results. Additionally, components that hit the force threshold of the UTM when $\varepsilon < 0.3$ were excluded due to the high probability that $\sigma_t^*$ was greater than the UTM's force threshold.

## 5.7 Maintenance

At the beginning of the campaign and periodically thereafter, new filament rolls were loaded into the printers. After performing material characterization (Section 3), a series of calibration components were printed to tune the extrusion multiplier of the printer to the density and diameter of the filament. The target mass for the calibration component was 3.3 g. If the component was too heavy, the extrusion multiplier was decreased. If it was too light, the extrusion multiplier was increased. This continued until the mass was within 5% of the target mass. In this way, it was possible to estimate the ratio of the filament length computed by Sli3er to the mass of the resulting component. As components were subsequently printed during the campaign, this ratio was slowly adjusted using integral tuning to remain accurate. Additionally, components were printed on polyimide tape that was applied to the glass bed of the printers. Whenever the tape showed signs of wear or became damaged, it was manually replaced.

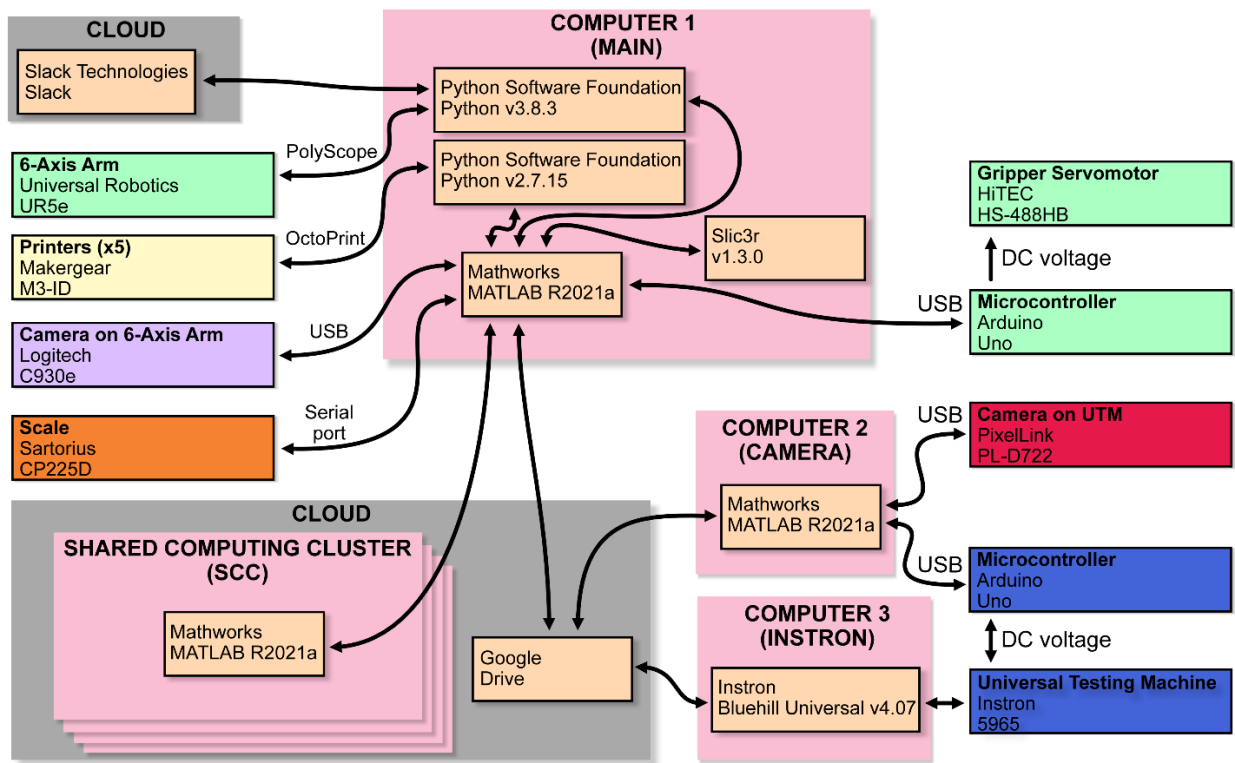

**Fig. S7 | Hardware and software Organization of the Bayesian experimental autonomous researcher (BEAR).**

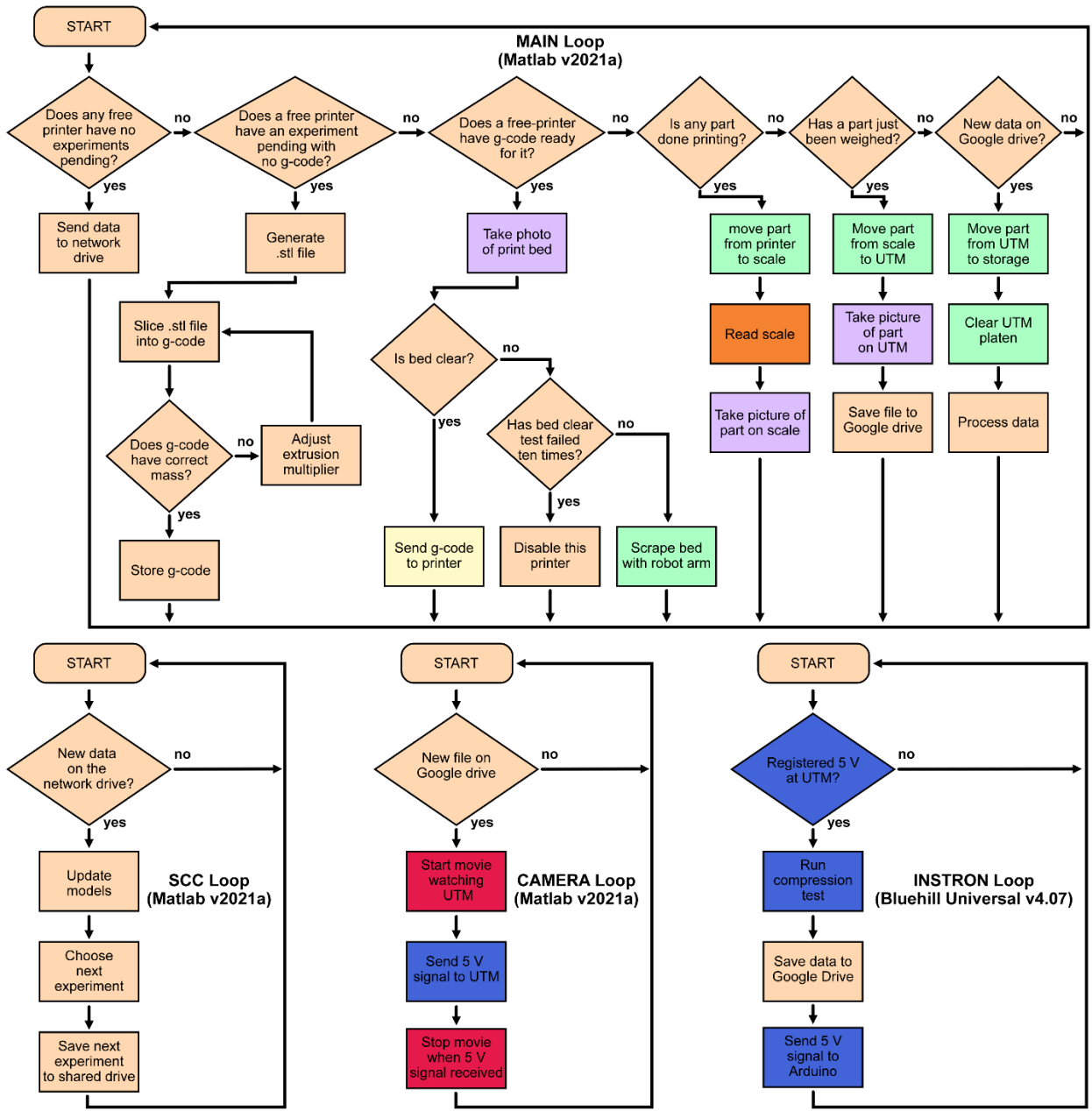

**Fig. S8 | Flowchart of the software loops executed in the four computers running as part of the BEAR.** Colors on the panels correspond to systems in Figure S7. Order of Main Loop actions can be adjusted by researchers to maximize throughput by prioritizing potential bottlenecks.

## 6. Details of the Experimental Campaign

The experimental campaign consisted of 25,387 experiments. During the course of the campaign, the available search space was changed by adding parameters, changing the limits of the included parameters, and by changing the method used to sample search space. After each experiment, the results were evaluated for defects in fabrication or testing. Components were excluded from the database if their height or mass deviated more than 5% from the target or if the maximum strain recorded was less than 30%. Additionally, researchers excluded components with severe print defects, which were reviewed daily. A record of all experiments performed is provided in Figure S9 and the raw data associated with these experiments is shared via kablab.org/data.

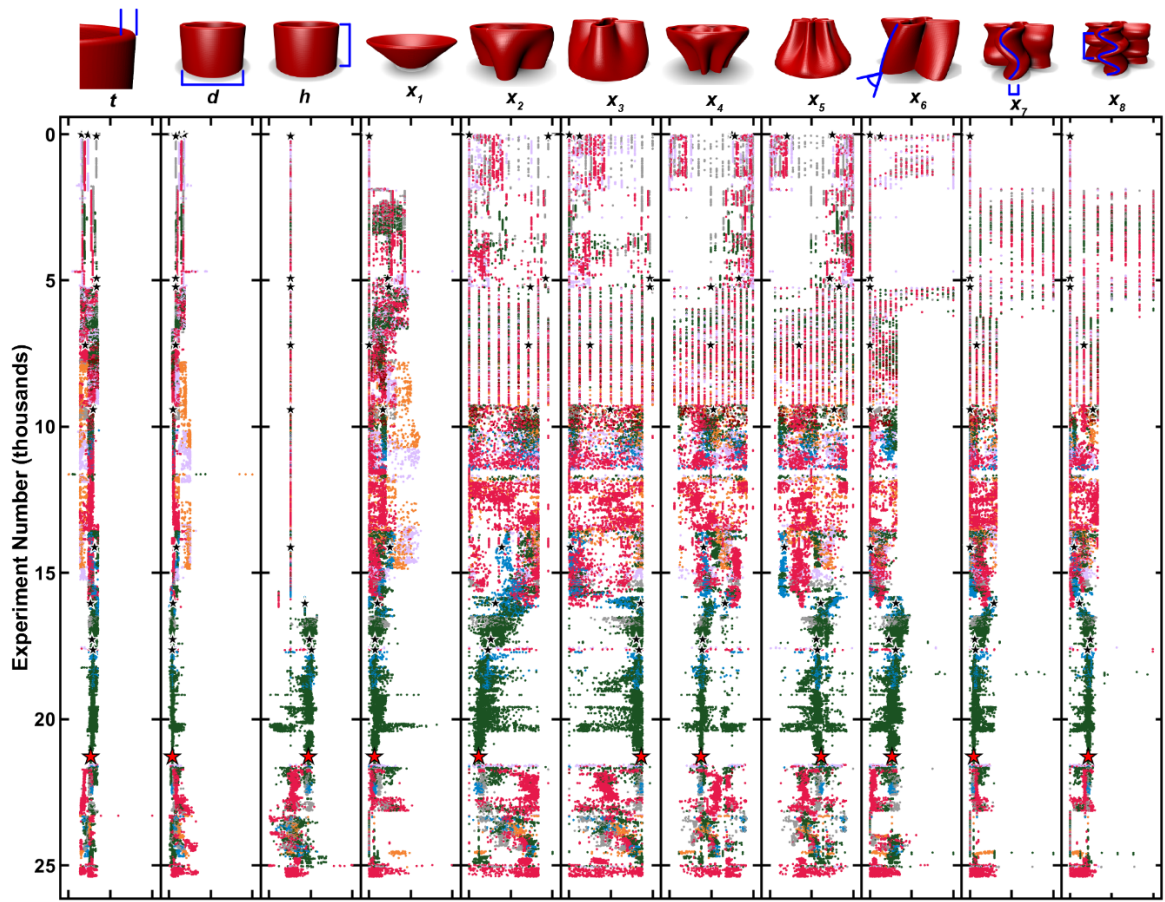

**Fig. S9 | Experiments carried out by the BEAR.** Experiments performed during the campaign, which are defined by eleven GCS parameter values. The color of each dot corresponds to the material used, as designated in Figure S4. Black and red stars correspond to breakthrough experiments, as designated in Figure 2a.

Over the course of the multi-year campaign, the details of how experiments were chosen were altered based on the intuition of the experimenters and by evaluating the progress of the BEAR. Examples of these changes include, the introduction of sinusoidal twist, the switch to LHS sampling (from grid-based sampling), allowing components to have both sinusoidal and linear twists combined, and switching to cooling plastic materials after printing. The timing of these changes is shown in Figure S10a. Researchers also controlled which filaments were loaded into which nozzles. New filaments were introduced during the campaign and the mix of filaments was changed to pursue different goals, as summarized in Figure S10b. Finally, 23 different decision policies were used throughout the campaign, as shown in Figure S10c and Table S3. Of particular importance was the introduction of $K_s^*$ as a key metric in decision policy six and the introduction of GPRs created by zooming in on the region of the best component to date, introduced with decision policy 18.

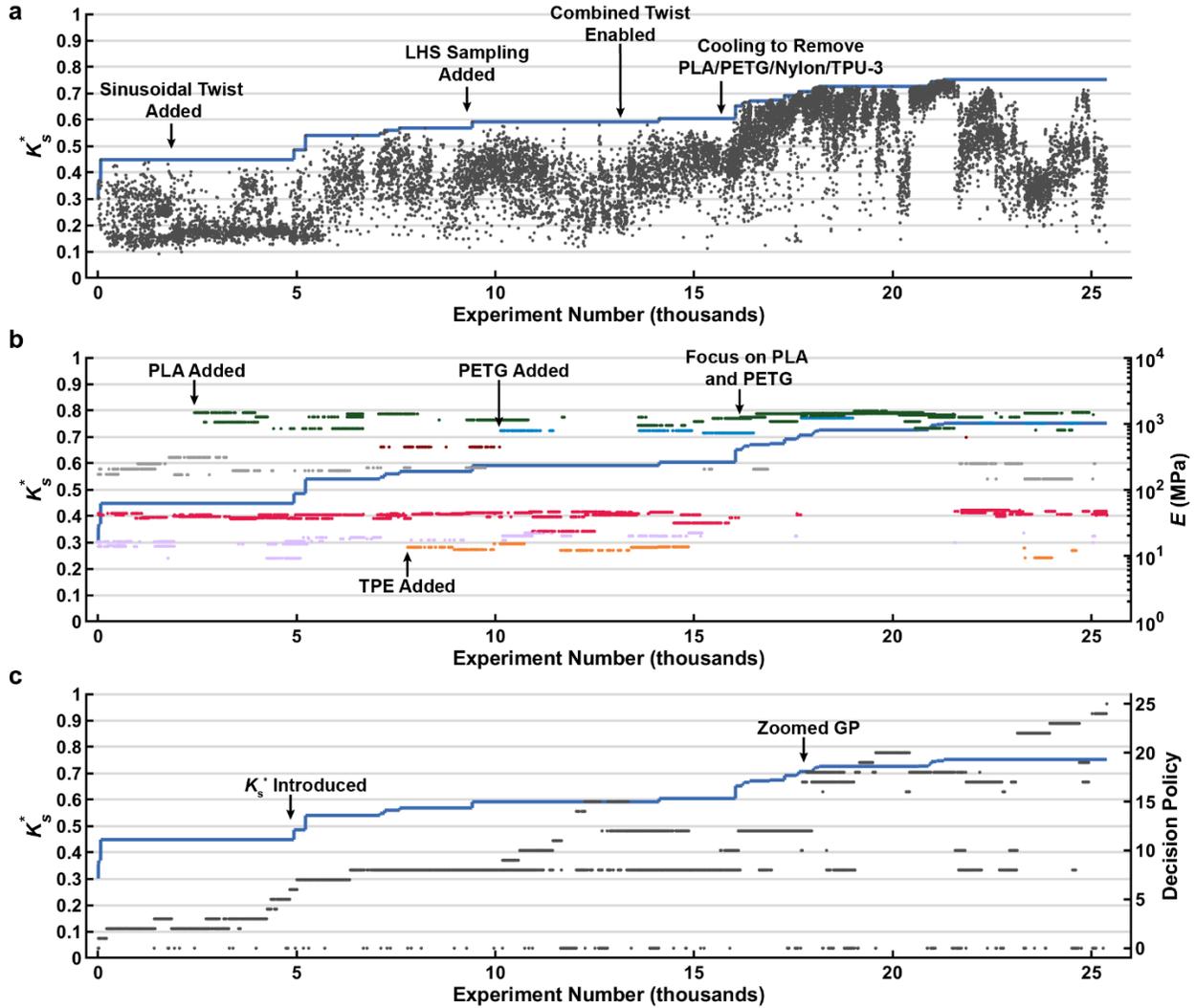

**Fig. S10 | Details of the human/machine collaboration. a**, $K_s^*$ of each successful test (gray dots), along with the highest $K_s^*$ to date (blue line). Key changes to the processing and sampling space are marked. **b**, Modulus of each experiment's filament roll plotted in semi-log (right axis) and colored according to the Figure S4, along with the highest $K_s^*$ to date (blue line – left axis). **c**, Decision policy of each experiment (right axis), along with the highest $K_s^*$ to date (blue line – left axis). Decision policies are listed with descriptions in Table S3.

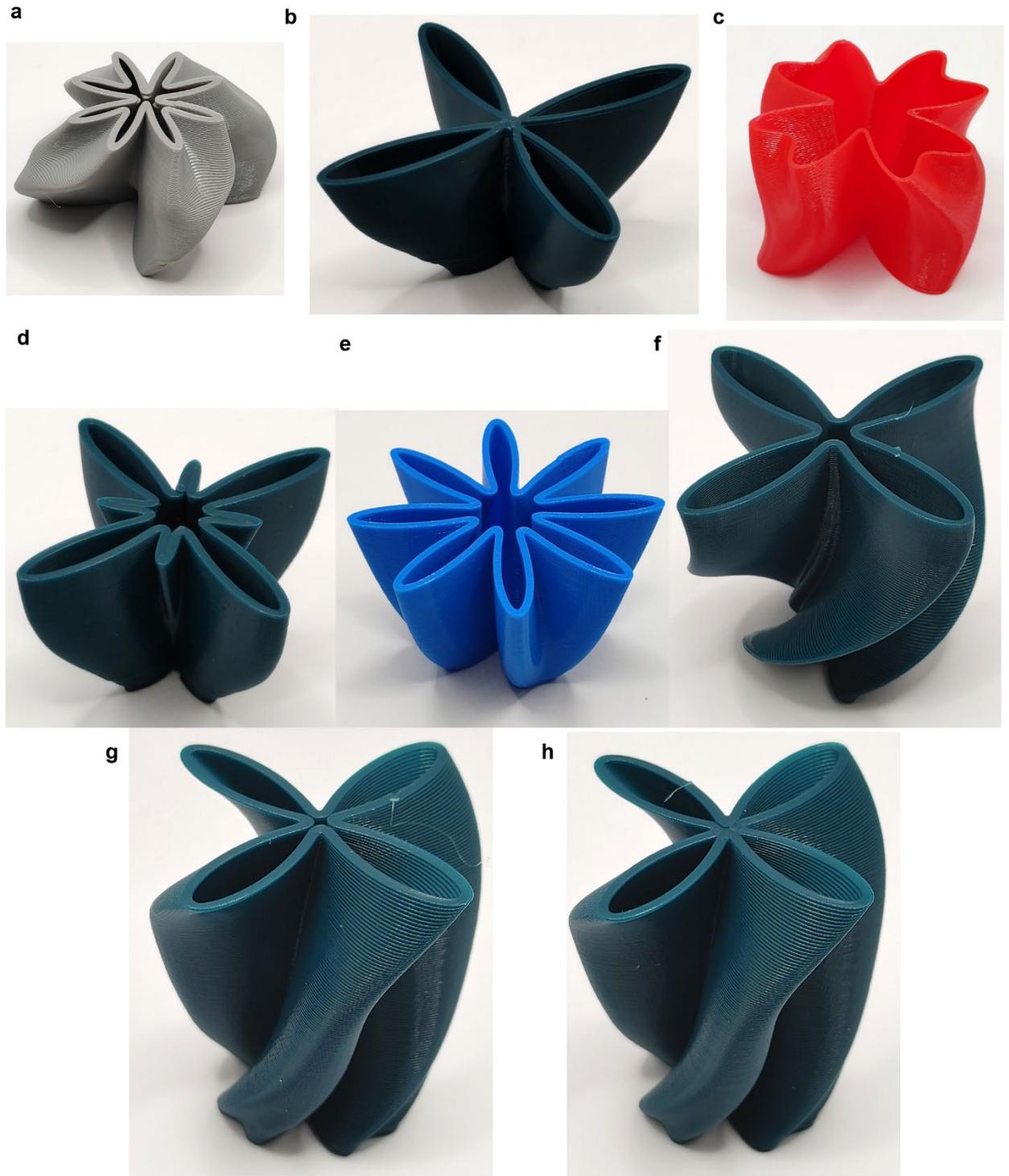

**Fig. S11 | Enlarged images of noteworthy components. a-h**, Pictures of noteworthy parts that significantly improved $K_s^*$ during the course of the campaign, as seen in Figure 2a. Heights vary from 19 mm (a-d) to 27.8 mm (g). Maximum widths range from 29 mm (g) to 48 mm (b). The color of the pictured components is indicative of the material used, with Green indicating PLA, Blue indicating PETG, Red indicating TPU-2, and Gray indicating TPU-3. Pictures are reprints, as the original parts were deformed during initial testing.

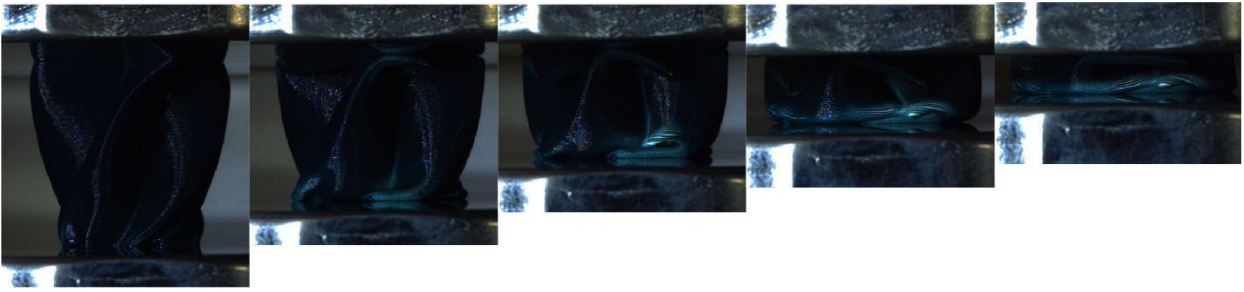

**Fig. S12 | As-recorded and uncorrected images for experiment 21,285.** Still images obtained from the video of this component being tested in uniaxial compression. Enhanced images and their corresponding strain are shown in Figure 2b.

## 7. Post-Analysis of the Campaign

Following the conclusion of the campaign, we sought to use the corpus of test results to understand the performance of superlative designs (Section 7.1), the performance envelopes of each material (Section 7.2), and the use of game theory to tease out the parameters responsible for the performance of the most efficient components (Section 7.3).

### 7.1 Comparison of the Iroko and Willow Designs

The superlative designs discovered in hyperelastic and plastic materials were very different. Plastic materials, which deform permanently, were able to achieve $K_s^* > 75\%$. Components made from hyperelastic materials, in contrast, were all $K_s^* < 63\%$ with consistent values being significantly lower still. Performance for superlative components made using the same design but different materials was correlated within material classes, but decreased significantly when moving outside the material class. For the top performing PLA design (Willow) and the top performing TPU-2 design (Iroko), three samples were printed on each of the five printers, for a total of 15 samples. Additionally, three samples were printed on a single printer for four other materials. All of these $\sigma$–$\varepsilon$ curves are shown in Figure S13.

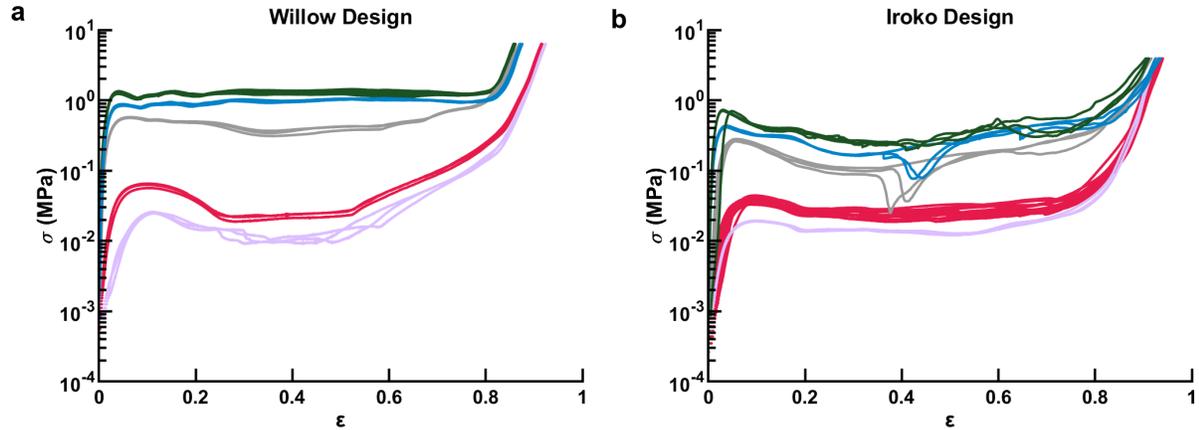

**Fig. S13 | Willow/Iroko by material. a**, Stress $\sigma$–strain $\varepsilon$ curves for components made using the Willow design printed in TPU-1, TPU-2, TPU-3, PETG, and PLA. PLA, the original Willow material, has 15 tests, while the other materials have three each. **b**, $\sigma$–$\varepsilon$ curves for components made using the Iroko design printed in the same five materials. TPU-2, the original Iroko material, has 15 tests while the other materials have three each. Colors depict the material as in Figure S4.

## 7.2 Material-Dependent Performance Envelope

The attainable envelope of $K_s^*$ and $\sigma_t^*$ for each material was estimated by computing a convex hull around all experimentally measured points (Figure S14). To determine the maximum stress $\sigma_{tp}$ for each material, the point with the highest $K_s^*$ was chosen. To obtain a measure of the uncertainty in this term, we retroactively step through the campaign and determine each time the $\sigma_{tp}$ would change and report the expected value as the median of these terms with the error being the standard deviation in their values (in logarithmic space).

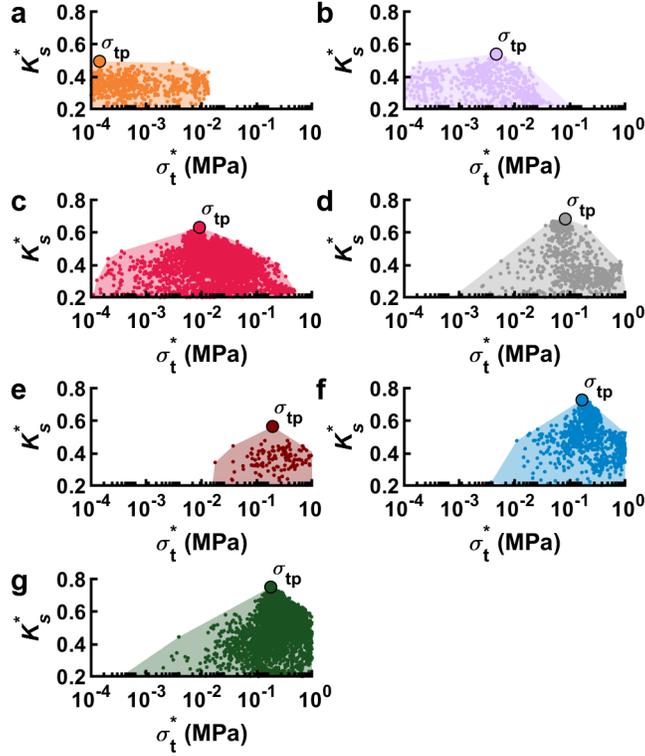

**Fig. S14 | Convex hulls for seven materials.** All tests for each of the seven materials studied, with their final $\sigma_{tp}$ marked. The materials are TPE (a), TPU-1-3 (b-d), nylon (e), PETG (f), and PLA (g).

### 7.3 Statistical Analysis of the Superlative Design

To understand how GCS parameters influence superlative components, we employed a machine learning-based approach to assess the significance of these parameters. Specifically, we built a neural network to predict $K_s^*$ for components made out of PLA. Figure S15a depicts the parity plot of this network. By applying Shapley additive explanations (SHAP),[4] we were able to separate the individual contributions made by each GCS parameter to the neural network's predictions of $K_s^*$. Inspired by Shapley values in Game Theory, SHAP assigns a value to each feature in a machine learning model, indicating its impact on the prediction. We seek to understand the difference in influence between a component and an ideal cylindrical shell (same diameter, height, and wall thickness). To achieve this, we subtract the SHAP values of Willow from the SHAP values for a pure cylindrical shell to obtain a "delta" in explanations. Our analysis of Willow revealed that the four most influential parameters contributing to its predicted performance are the wavelength of the sinusoidal twist ($x_8$), the linear twist linear ($x_6$), and the 4-period amplitude of the bottom and top ($x_2$, $x_3$), (Figure S15b).

The neural network used for SHAP analysis comprised six layers: a 64-dimension linear layer followed by a ReLU activation[5], repeated three times. A data split of 80% for training, 10% for validation, and 10% for testing was employed. The GCS parameters were normalized and no preprocessing was applied to $K_s^*$. The network was trained using the mean squared error loss function. The training process uses the Adam optimizer[6] with a learning rate of 0.001, weight decay of $1\times10^{-5}$, and a batch size of 16. Training was performed for 500 epochs with early stopping. The PLA network achieved a test loss of 0.0032 and a coefficient of determination $R^2 = 0.88$. For interpreting the predictions generated by the neural networks, we used the SHAP DeepExplainer which is initialized using the training split data. To provide explanations for individual components, we use the default SHAP waterfall visualization.

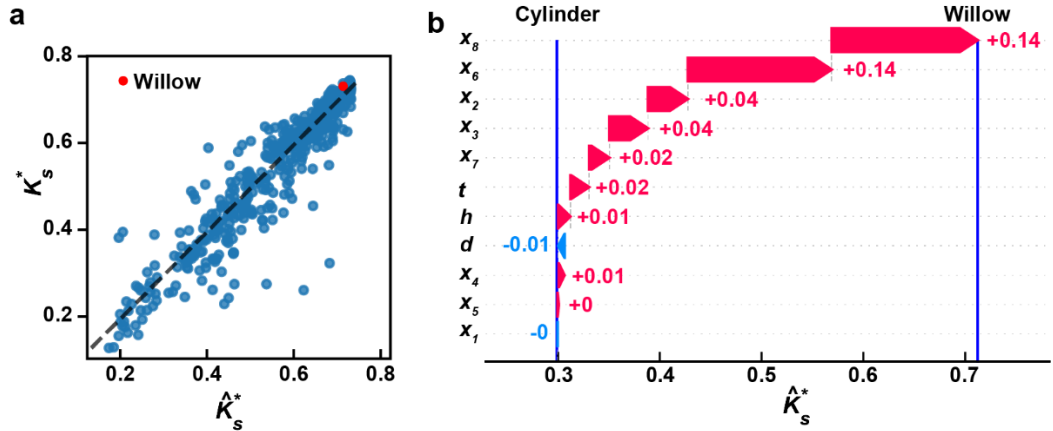

**Fig. S15 | Shapley additive explanations (SHAP) analysis of Willow as the superlative design.** **a**, Parity plot of the neural network built on all data taken using PLA with Willow highlighted. **b**, SHAP waterfall plot for the Willow design tested in PLA relative to a PLA cylindrical shell with the same height, diameter, and thickness. These values show the cumulative effect of positive (red) or negative (blue) contributions of individual feature values to model predictions.

## 8. Supplemental References